\documentclass[english,11pt]{article}
\usepackage{epsf,amsmath,amssymb,graphicx,dcolumn}
\usepackage{scalefnt}
\usepackage{subfigure,ulem}
\usepackage{pstricks}
\usepackage{cite,hyperref}
\usepackage{color}
\usepackage{rotating}
\usepackage{feynmp}
\newcommand{\pdf}{{\abbrev PDF}}
\newcommand{\qcd}{{\abbrev QCD}}

\newcommand{\abbrev}{\scalefont{.9}}

\newcommand{\ptH}{p_\text{T,H}}

\newcommand{\muF}{\mu_\text{F}}
\newcommand{\muR}{\mu_\text{R}}
\newcommand{\mhiggs}{\MH}
\newcommand{\mtop}{\Mt}
\newcommand{\mbottom}{\Mb}
\newcommand{\ep}{\epsilon}
\newcommand{\api}{\frac{\alphas}{\pi}}
\newcommand{\eqn}[1]{Eq.\,(\ref{#1})}

\newcommand{\dd}{{\rm d}}
\newcommand{\deriv}[2]{\frac{\dd #1}{\dd #2}}
\newcommand{\order}[1]{{\cal O}(#1)}
\newcommand{\lhc}{{\abbrev LHC}}
\newcommand{\sm}{{\abbrev SM}}

\newcommand{\lo}{\text{\abbrev LO}}
\newcommand{\nlo}[1]{\text{\abbrev N$^{#1}$LO}}
\newcommand{\nnlo}{\text{\abbrev NNLO}}

\newcommand{\msbar}{\overline{\mbox{\abbrev MS}}}

\renewcommand{\Re}{{\rm Re}}

\def\lra{\mathop{\mathrm{\leftrightarrow}}\nolimits}

\def\asymp#1%
{\mathrel{\raisebox{-.4em}{$\widetilde{\scriptstyle #1}$}}}

\def\Nequal#1%
{\mathrel{\raisebox{-.5em}{$\stackrel{=}{\scriptstyle\rm#1}$}}}
\newcommand{\dsl}[1]{\not \hspace{-0.7mm}#1}
\def\dsl{\mathpalette\make@slash}
\def\make@slash#1#2{\setbox\z@\hbox{$#1#2$}%
  \hbox to 0pt{\hss$#1/$\hss\kern-\wd0}\box0}

\def\beq{\begin{equation}}
\def\eeq{\end{equation}}
\def\bit{\begin{itemize}}
\def\eit{\end{itemize}}
\def\beqar{\begin{eqnarray}}
\def\eeqar{\end{eqnarray}}
\def\barr#1{\begin{array}{#1}}
\def\earr{\end{array}}
\def\bfi{\begin{figure}}
\def\efi{\end{figure}}
\def\btab{\begin{table}}
\def\etab{\end{table}}
\def\bce{\begin{center}}
\def\ece{\end{center}}
\def\nn{\nonumber}

\def\al{\alpha}

\def\ga{\gamma}

\def\eps{\epsilon}
\def\veps{\varepsilon}

\def\la{\lambda}

\def\si{\sigma}

\def\refeq#1{\mbox{Eq.\,(\ref{#1})}}

\def\reffi#1{\mbox{Fig.\,\ref{#1}}}

\def\refta#1{\mbox{Table\,\ref{#1}}}

\def\refse#1{\mbox{Section\,\ref{#1}}}

\def\citere#1{\mbox{Ref.\cite{#1}}}
\def\citeres#1{\mbox{Refs.\cite{#1}}}

\newcommand{\TeV}{\unskip\,\mathrm{TeV}}
\newcommand{\GeV}{\unskip\,\mathrm{GeV}}

\newcommand{\ri}{{\mathrm{i}}}

\newcommand{\M}{{\cal{M}}}

\def\mathswitchr#1{\relax\ifmmode{\mathrm{#1}}\else$\mathrm{#1}$\fi}

\newcommand{\PW}{\mathswitchr W}

\newcommand{\PZ}{\mathswitchr Z}

\newcommand{\Pg}{\mathswitchr g}

\newcommand{\PH}{\mathswitchr H}
\newcommand{\PG}{\mathswitchr G^0}

\newcommand{\Pb}{\mathswitchr b}

\newcommand{\Pp}{\mathswitchr p}
\newcommand{\Pt}{\mathswitchr t}

\newcommand{\Pq}{q}

\def\mathswitch#1{\relax\ifmmode#1\else$#1$\fi}

\newcommand{\MW}{\mathswitch {M_\PW}}

\newcommand{\MZ}{\mathswitch {M_\PZ}}
\newcommand{\MH}{\mathswitch {M_\PH}}

\newcommand{\Mb}{\mathswitch {m_\Pb}}
\newcommand{\Mt}{\mathswitch {m_\Pt}}

\newcommand{\rw}{\mathswitchr w}
\newcommand{\sw}{\mathswitch {s_\rw}}
\newcommand{\cw}{\mathswitch {c_\rw}}

\newcommand{\alphas}{\al_{\mathrm{s}}}

\makeatletter

\def\paragraph{\@startsection{paragraph}{4}{\z@}{+2.00ex plus
 +1ex minus +.2ex}{1.5ex plus .2ex}{\it\normalsize}}

\makeatother

\def\draftdate{\relax}

\def\mda{\relax}
\def\mua{\relax}
\def\mla{\relax}
\def\Mda{\relax}
\def\Mua{\relax}
\def\Mla{\relax}
\def\draft{
\def\thtystars{******************************}
\def\sixtystars{\thtystars\thtystars}
\typeout{}
\typeout{\sixtystars**}
\typeout{* Draft mode!
         For final version remove \protect\draft\space in source file *}
\typeout{\sixtystars**}
\typeout{}
\def\draftdate{\today}
\def\mua{\marginpar[\boldmath\hfil$\uparrow$]%
                   {\boldmath$\uparrow$\hfil}%
                    \typeout{marginpar: $\uparrow$}\ignorespaces}
\def\mda{\marginpar[\boldmath\hfil$\downarrow$]%
                   {\boldmath$\downarrow$\hfil}%
                    \typeout{marginpar: $\downarrow$}\ignorespaces}
\def\mla{\marginpar[\boldmath\hfil$\rightarrow$]%
                   {\boldmath$\leftarrow $\hfil}%
                   \typeout{marginpar: $\lra$}\ignorespaces}
\def\Mua{\marginpar[\boldmath\hfil$\Uparrow$]%
                   {\boldmath$\Uparrow$\hfil}%
                    \typeout{marginpar: $\uparrow$}\ignorespaces}
\def\Mda{\marginpar[\boldmath\hfil$\Downarrow$]%
                   {\boldmath$\Downarrow$\hfil}%
                    \typeout{marginpar: $\downarrow$}\ignorespaces}
\def\Mla{\marginpar[\boldmath\hfil$\Rightarrow$]%
                   {\boldmath$\Leftarrow $\hfil}%
                    \typeout{marginpar: $\lra$}\ignorespaces}
\overfullrule 5pt
\oddsidemargin -15mm
\oddsidemargin -10mm
\marginparwidth 29mm
}

\title{\vspace*{-6em}
  \begin{flushright}
    {\small 
       CERN-PH-TH/2012-312\\
       FR-PHENO-2012-023 \\[-1em]
       WUB/12-21
    }
  \end{flushright}
\vspace*{2em} {\bf Gluon-induced Higgs-strahlung \\ at next-to-leading
  order QCD}}
\author{Lukas Altenkamp$^{a}$, Stefan Dittmaier$^a$,\\ 
Robert V. Harlander$^b$, Heidi Rzehak$^{a,c}$, Tom J.E.~Zirke$^b$\\[2em]
$^a$ {\it Physikalisches Institut, Albert-Ludwigs-Universit\"at Freiburg,}\\
{\it  D-79104 Freiburg, Germany}\\[.3em]
$^b$ {\it Fachbereich C,
  Bergische Universit\"at Wuppertal,}\\[0em] {\it 42097 Wuppertal,
  Germany}\\[.3em]
$^c$ {\it TH Division, Physics Department, CERN}\\[0em]
{\it CH-1211 Geneva 23, Switzerland}
}
\date{}
\begin{document}
\maketitle

\vspace*{1cm}
\begin{abstract}
Gluon-induced contributions to the associated production of a Higgs and
a $\PZ$~boson are calculated with \nlo{} accuracy in \qcd{}. They
constitute a significant contribution to the cross section for
this process. The perturbative correction factor ($K$-factor) is
calculated in the limit of infinite top-quark and vanishing bottom-quark
masses. The qualitative similarity of the results to the well-known ones
for the gluon-fusion process $\Pg\Pg\to\PH$ allows to conclude that rescaling
the \lo{} prediction by this $K$-factor leads to a reliable \nlo{}
result and realistic error estimate due to missing
higher-order perturbative
effects. We consider the total inclusive cross section as well as
a scenario with a boosted Higgs boson,
where the Higgs boson's transverse momentum is restricted to
values $\ptH{}>200\GeV$. In both cases, we find large correction factors
$K\approx 2$ in most of the parameter space.
\end{abstract}

\vfill

November 2012
\thispagestyle{empty}

\clearpage

\section{Introduction}

With the recent observation
of a new particle at the
\lhc{}~\cite{Aad:2012gk,Chatrchyan:2012gu} and the related evidence at the
Tevatron~\cite{Aaltonen:2012qt}, efforts to determine its identity are
of highest priority. Among the most important observables are the total
and differential cross sections. First measurements of these quantities
indicate that the new particle is indeed the long-sought Higgs boson of
the Standard Model (\sm). In order to definitely confirm or exclude this
hypothesis, accurate measurements and corresponding precision
calculations of the cross section in the various production modes are
required.

The current theoretical knowledge of the \sm{} cross sections is in general
quite impressive and documented in
\citeres{Dittmaier:2011ti,Dittmaier:2012vm}. Subject of the current
paper is a particular contribution to the so-called Higgs-strahlung
process $\Pp\Pp\to\PH V$ ($V=\PW,\PZ$). While it has been
a major
search mode for Higgs bosons at the Tevatron, it used to be considered
of minor importance at the \lhc{} due to its small cross section and
large background. However, it belongs to the channels that were analysed
by the {\abbrev ATLAS} and the {\abbrev CMS} experiments already with
the first data.  The signal-to-background ratio for Higgs-strahlung can
be significantly enhanced when cutting on events where the Higgs boson
is produced at large $\ptH$~\cite{Butterworth:2008iy}.

The leading-order (\lo) cross section for this process can be written as a convolution
of the cross section for the Drell--Yan process $\Pp\Pp\to V^\ast$ with the
decay rate for $V^\ast\to\PH V$, where $V^\ast$ denotes an off-shell gauge
boson of momentum $k$:
\begin{equation}
\begin{split}
\sigma^\text{HV,DY}(\Pp\Pp\to \PH V) = \int\dd k^2\,\sigma^\text{DY}(\Pp\Pp\to
V^\ast)\,\deriv{\Gamma(V^*\to\PH V)}{k^2}.
\label{eq:sigdy}
\end{split}
\end{equation}
This relation holds exactly through next-to-leading order (\nlo{})
\qcd{}, i.e.\ $\order{\alphas}$, and approximately through
next-to-next-to-leading order (\nnlo).  The \qcd{} effects of
\refeq{eq:sigdy} are therefore strongly dominated by the Drell--Yan
corrections to the cross section $\sigma^\text{DY}$; they are known
through \nnlo{} \qcd{} for the total $\PH\PW/\PH\PZ$ cross
sections~\cite{Hamberg:1991np,Harlander:2002wh,Brein:2003wg}, and for
$\PH\PW$ production also differentially~\cite{Ferrera:2011bk}.  Typical
Feynman diagrams for the Drell--Yan type contribution are shown in
\reffi{fig:diasa}\,(a-f).  They contribute to the cross section at order
$g^4\alphas^n$ ($n=0,1,2$) and increase it by about $30\%$ with respect
to \lo. Here and in what follows, $\alphas=g_s^2/(4\pi)$, with $g_s$ the
strong and $g$ the weak coupling constant.

Apart from the Drell--Yan-like 
\qcd{} corrections at \nnlo{}, there are
top-loop-induced contributions such as the ones shown in
\reffi{fig:diasa}\,(g-j). Their interference with the \lo{} and the
real-emission \nlo{} amplitude is of order $\lambda_{\Pt}g^3\alphas^2$,
with $\lambda_{\Pt}$ the top Yukawa coupling, and their numerical impact
is at the percent level~\cite{Brein:2011vx}.

In contrast to the \nlo{} \qcd{} and dominant \nnlo{} \qcd{}
corrections, electroweak ({\abbrev EW}) corrections do not respect a
factorization into Drell--Yan-like production and decay, since
irreducible (box) corrections to $\Pq\Pq^{(\prime)}\to\PH V$ already
contribute at \nlo{}.  The \nlo{} {\abbrev EW} corrections have been
evaluated in \citere{Ciccolini:2003jy} for the total $\PH V$ cross
sections, where they amount to $-(5{-}10)\%$, and in
\citere{Denner:2011id} 
for differential distributions as part of the {\tt HAWK} Monte Carlo
program, which fully includes all decays and off-shell effects of the
weak boson $V=\PW,\PZ$. In distributions the {\abbrev EW} corrections
can grow to $-(10{-}20)\%$.  As suggested in \citere{Brein:2004ue},
\nlo{} {\abbrev EW} and Drell--Yan-like \nnlo{} \qcd{} corrections can
be conveniently combined in factorized form, where the {\abbrev EW}
corrections modify the \qcd{} prediction by a relative correction factor
that is rather insensitive to the parton luminosities.

Recently, \qcd{} corrections to the $\PH\to\Pb\bar\Pb$ decay have been
considered as well~\cite{Banfi:2012jh}. These final-state corrections
should be carefully taken into account in the Higgs reconstruction.

In this paper we focus on another type of contribution which is specific
to $\PH\PZ$ production, namely gluon fusion, mediated by top- and
bottom-quark loops.  Typical diagrams of this channel are shown in
\reffi{fig:diasb}. Owing to the initial-state gluons, it cannot
interfere with the \lo{} amplitude and therefore contributes to the
cross section at order $\lambda_{\Pt}^2g^2\alphas^2$.  For
$\mhiggs=125\GeV$, at leading, i.e., one-loop order it amounts to about
4\% (6\%) of the total Higgs-strahlung cross section at the \lhc{} with
$8\TeV$ ($14\TeV$)~\cite{Brein:2003wg}.  Since it has no lower-order
correspondence, it is separately gauge invariant and {\abbrev IR} and
{\abbrev UV} finite. The two initial-state gluons lead to a rather
strong renormalization and factorization scale dependence of about
$30\%$, thus increasing the theoretical uncertainty of the $\PH\PZ$
relative to the $\PH\PW$ process, where the $\Pg\Pg$ channel does not
exist (at this order).  Experience from the gluon-fusion process
$\Pg\Pg\to \PH$ shows, however, that the \lo{} scale uncertainty
drastically underestimates the actual size of the higher-order
corrections. Owing to the similarity of the $\Pg\Pg\to \PH$ and the
$\Pg\Pg\to \PH\PZ$ processes in their \qcd{} structure (same initial
states and colour structure, both loop-induced), we expect a similar
phenomenon in the latter.

The goal of the present paper is to improve on the theory uncertainty of
the $\Pg\Pg\to \PH\PZ$ process by calculating its \nlo{} \qcd{}
corrections. Note that they are of order $\alphas^3$ and thus {\it
  formally} contribute to the \nlo{3} corrections of the Higgs-strahlung
process.
Technically the described \nlo{} calculation involves massive, multi-scale
two-loop diagrams that are beyond present calculational techniques, so that
we are forced to employ asymptotic expansions in the limit of a large
top-quark mass.
We note that the same strategy was already successfully applied to the
calculation of \nlo{} corrections to the
related process of Higgs pair production via gluon fusion,
$\Pg\Pg\to\PH\PH$~\cite{Dawson:1998py}.

The paper is organized as follows:
In \refse{sec::outline} we briefly outline the problem, before describing the
details of our calculation in \refse{sec:details}.
Our numerical results are discussed in \refse{sec:numerics}, and our
conclusions given in \refse{sec::conclusions}.

\section{Outline of the problem}\label{sec::outline}
\label{sec:outline}

\subsection{Leading order}

At \lo{} and in covariant $R_\xi$ gauge, the Feynman diagrams
contributing to the gluon-induced Higgs-strahlung process can be divided
into three types, shown in \reffi{fig:diasb}:
\begin{figure}
\begin{center}
\includegraphics[width=\textwidth,trim=4.4cm 12cm 4.4cm
4.4cm,clip]{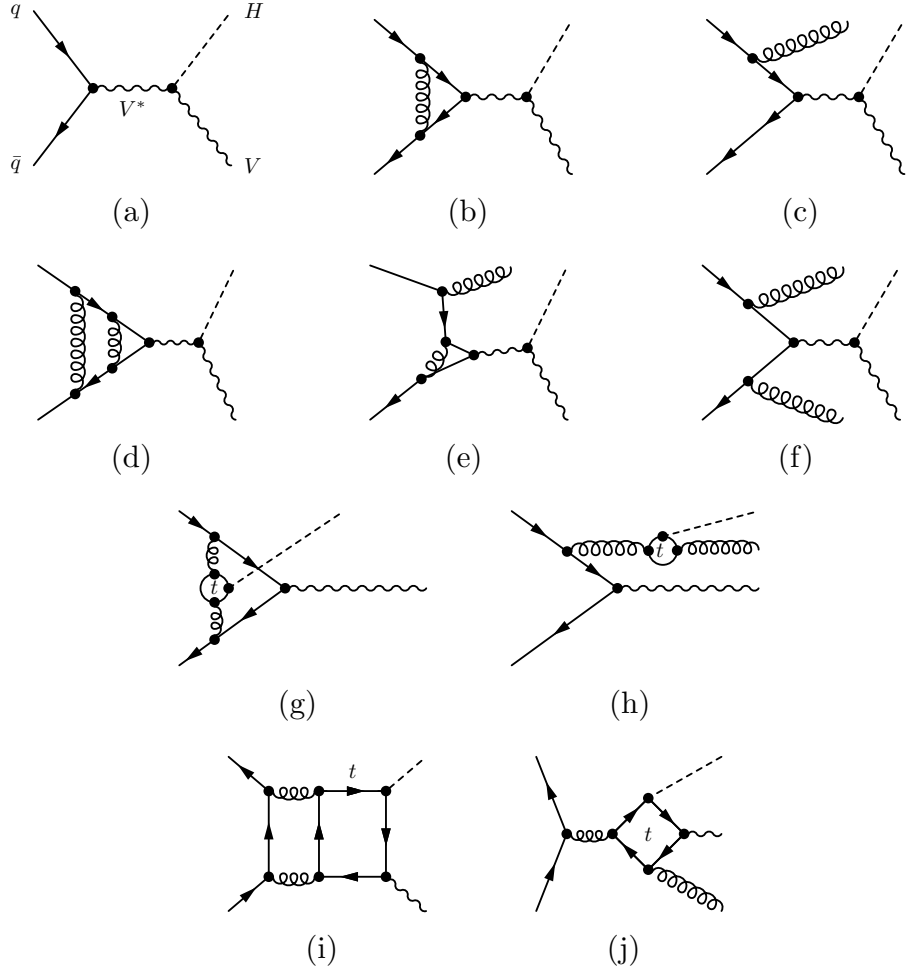}
\caption[]{\label{fig:diasa} Representative diagrams to hadronic
  $\PH\PZ$ production of Drell--Yan type up to \nnlo{} (a-f) and
  non-Drell--Yan-like \nnlo{} graphs with Higgs radiation off top-quark
  loops; both types of corrections (up to \nnlo{}) are {\it not} considered
  in this publication.}
\end{center}
\end{figure}
\renewcommand{\labelenumi}{(\alph{enumi})}
\begin{enumerate}
\item Box diagrams for $\Pg\Pg\to\PH\PZ$: Only massive quarks run in the
  loop due to the proportionality to the respective Yukawa
  coupling. Note, however, that these graphs tend to zero also in the
  heavy-quark limit.
\item Triangle diagrams for $\Pg\Pg\to \PZ^\ast\to\PH\PZ$: Owing to
  Furry's theorem, all contributions from vector couplings compensate
  each other, so that only the axial-vector coupling of the $\PZ$~boson
  needs to be taken into account. Since the axial-vector coupling is
  proportional to the third component of the weak isospin of the quark
  ($\pm\frac{1}{2}$), the contribution of a single quark generation
  vanishes in the equal-mass case.  Assuming massless quarks in the
  first two fermion generations, this leaves a non-vanishing
  contribution only from the third generation.  The amplitudes tend to
  zero in the heavy-quark limit.
  
  It is interesting to note that only the longitudinal part of the
  $\PZ$-boson propagator contributes, while all contributions of the
  transverse part vanish.  This consequence of the Landau--Yang 
  theorem~\cite{landau,Yang:1950rg}
  can be used at \nlo{} to facilitate the calculation significantly, as
  will be described below.
\item Triangle diagrams for $\Pg\Pg\to \PG\to\PH\PZ$: Only the massive-quark 
  loops contribute here, where $\PG$ is the would-be Goldstone boson partner
  to the $\PZ$~boson. The graphs are both proportional to the respective Yukawa
  coupling and to the third component of the weak isospin of the quark and
  tend to a constant in the heavy-quark limit.
\end{enumerate}
While the box diagrams (a) are gauge-parameter independent in the
$R_\xi$ gauge, both the vertices (b) and (c) depend on the
gauge-parameter of the $\PZ$~boson.  The sum of (b) and (c)
for each quark generation is, of course, gauge-parameter independent.  

The full result for the \lo{} amplitudes for the process $\Pg\Pg\to \PH\PZ$
can be found in \citere{Kniehl:1990iv}; the hadronic cross section can
be easily obtained using the program {\tt
  vh@nnlo}~\cite{Brein:2003wg,vhnnlo}.  We have rederived the \lo{} cross
section with the full dependence on the top- and bottom-quark masses as
a basic ingredient of our \nlo{} calculation.

\subsection{Next-to-leading order}

The Feynman diagrams for the \nlo{} \qcd{} corrections to the
$\Pg\Pg\to\PH\PZ$ process are obtained from the \lo{} gluon-fusion
diagrams shown in \reffi{fig:diasb}\,(a-c) by attaching virtual and real
gluons and quarks to internal and external quark and gluon lines in all
possible ways:
\begin{figure}
\begin{center}
\includegraphics[width=\textwidth,trim=4.4cm 8cm 4.4cm
4.4cm,clip]{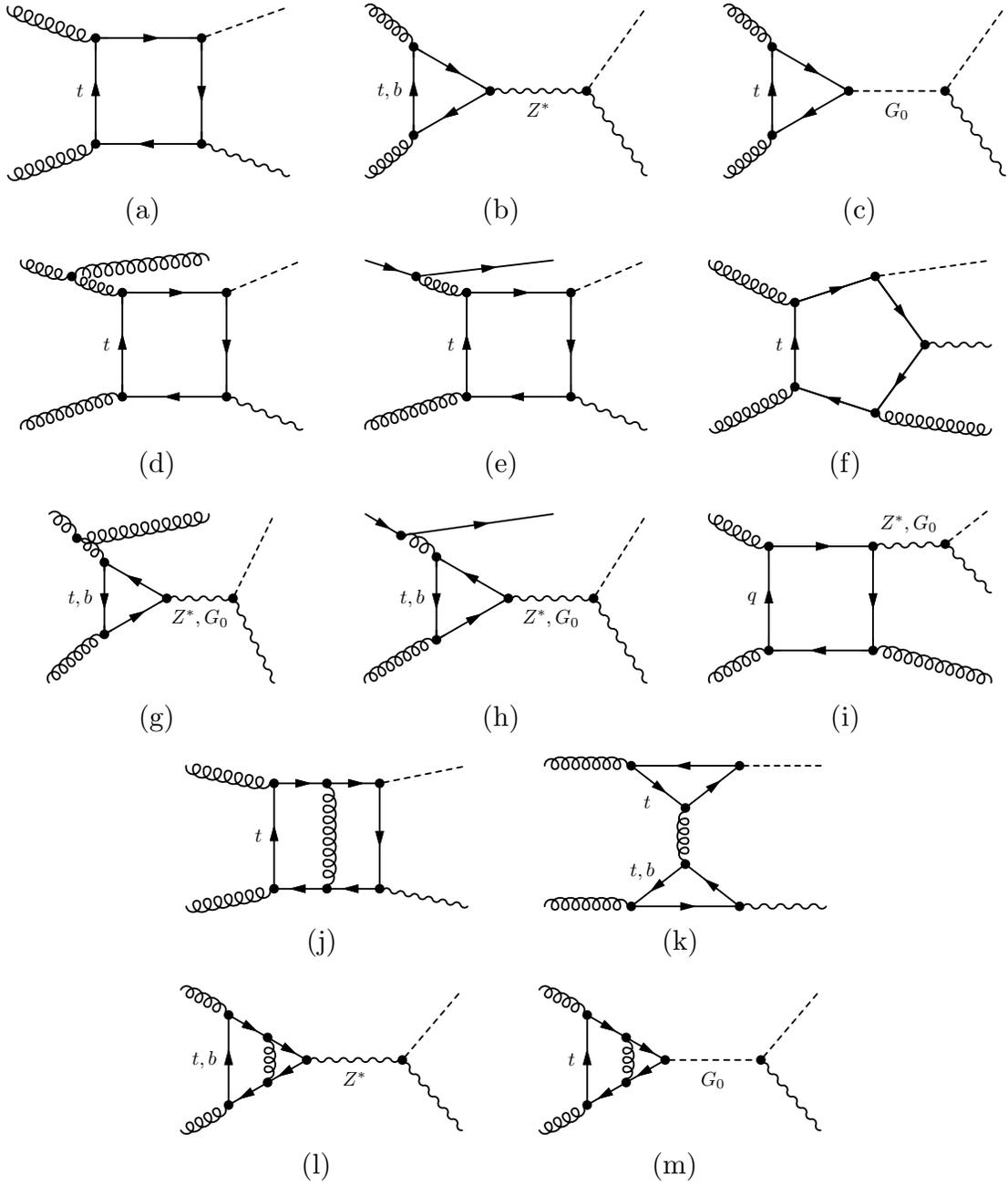}
\caption[]{\label{fig:diasb}%
Representative diagrams to hadronic $\PH\PZ$ production via quark-loop-induced
gluon fusion. It is understood that crossed diagrams have to be taken into account as well.}
\end{center}
\end{figure}
\begin{itemize}
\item
For the real corrections, this results in triangle diagrams with two
massive external momenta, \reffi{fig:diasb}\,(g,h), box diagrams with
one or three massive external momenta, \reffi{fig:diasb}\,(i) and (d,e),
and pentagon diagrams with two massive external momenta,
\reffi{fig:diasb}\,(f) (counting off-shell gluons as massive lines).
Note that \reffi{fig:diasb}\,(e) is a crossed version of
\reffi{fig:diasa}\,(j), for example; as pointed out above, it can
interfere at $\order{\lambda_{\Pt}g^3\alphas^2}$ with the \nlo{}
real-emission Drell--Yan type amplitude, which has already been taken
into account in \citere{Brein:2011vx}. In the present paper, we work at
$\order{\lambda_{\Pt}^2g^2\alphas^3}$ and need to evaluate the square of
such terms.
\item
\begin{sloppypar}
For the virtual corrections, we encounter two-loop vertex and box
diagrams, \reffi{fig:diasb}\,(j,l,m), as well as one-particle-reducible
diagrams with two one-loop triangle insertions, \reffi{fig:diasb}\,(k).
\end{sloppypar}
\end{itemize}

\section{Details of the calculation and effective-field-theory approach}
\label{sec:details}

While the majority of the integrals could be calculated using well-known
techniques, a general result for the massive double-box integrals shown
in \reffi{fig:diasb}\,(j) is beyond current technology. However,
motivated by observations made in
\citeres{Spira:1995rr,Kramer:1996iq,Harlander:2010my,Harlander:2009bw,%
  Pak:2009dg,Pak:2011hs,Harlander:2012hf}, for example, we follow a
strategy that has been successfully applied to higher-order corrections
to Higgs production via gluon fusion. Instead of calculating the Feynman
integrals in full generality, we determine the perturbative correction
factor
\begin{equation}
\begin{split}
K = \frac{\sigma^\nlo{}}{\sigma^\lo}
\label{eq:kfacdef}
\end{split}
\end{equation}
in the limit of infinite top-quark and vanishing bottom-quark masses (referred
to as ``effective theory'' in what follows). For the gluon-fusion
process, both inclusive and differential, it turns out that this factor
is rather insensitive to the top-quark mass
effects~\cite{Spira:1995rr,Kramer:1996iq,Harlander:2010my,Harlander:2009bw,%
  Pak:2009dg,Pak:2011hs,Harlander:2012hf}. Using asymptotic expansion of
Feynman diagrams~\cite{Smirnov:2002pj,Smirnov:1994tg}, the heavy-top
limit can be interchanged with the loop integration, which simplifies
the calculation enormously.

\paragraph{\lo{} amplitude}
At \lo{}, the diagrams with top-quark loops reduce to vacuum diagrams
(integrals with vanishing external momenta) in the effective theory.
Because already at \lo{} the loop integrals are {\abbrev UV} divergent,
some care is needed in the calculation of the Dirac traces that involve
the matrix $\ga_5$.  Both at \lo{} and \nlo{}, we consistently use the
't~Hooft--Veltman scheme~\cite{'tHooft:1972fi,Breitenlohner:1977hr},
where $\ga_5$ anticommutes with the first four, but commutes with all
other Dirac matrices.  In practice, we insert
$\ga_5=-\frac{\ri}{4!}\eps_{\mu\nu\rho\si}\ga^\mu\ga^\nu\ga^\rho\ga^\si$
($\epsilon^{0123}=+1$), 
keep the $\eps$-tensor outside of the
$D$-dimensional integration, and project onto four dimensions only after
all divergent terms have cancelled among each other.

The result for the \lo{} amplitude $\M_0$ and its polarization- and
colour-averaged square is
\mda
\begin{equation}
\begin{split}
\M_0 &=
-\frac{\alphas\alpha}{\sw^2\cw^2\MZ} \,\delta^{ab}
\eps(\veps_1,\veps_2,p_1,p_2) \,\frac{p_\PH\cdot\veps_\PZ^*}{\hat s},
\\ 
\overline{|\M_0|^2} &= \frac{\alphas^2\alpha^2}{256\sw^4\cw^4\MZ^2}
\, \lambda(\hat{s},\MH^2,\MZ^2), 
\label{eq:M0square} 
\end{split}
\end{equation}
with $\MH$ the Higgs mass and $\MZ$ the $\PZ$ mass,
$\alpha$ and $\alphas$ the electromagnetic and the strong coupling
constants, $\sw^2=1-\cw^2$ the sine of the weak mixing angle, and
$\lambda(a,b,c)=a^2+b^2+c^2-2ab-2ac-2bc$.  
In \eqn{eq:M0square}
\mua
we make use of the
momentum assignment $\Pg^a(p_1)+\Pg^b(p_2)\to\PH(p_\PH)+\PZ(p_\PZ)$,
with $a,b$ denoting colour indices and $\hat s=(p_1+p_2)^2$ the usual
Mandelstam variable.  The polarization vectors $\veps_i(p_i)$ ($i=1,2$)
correspond to the respective incoming gluons, and $\veps_\PZ^*(p_\PZ)$
to the outgoing $\PZ$~boson.  The shorthand
$\eps(\veps_1,\veps_2,p_1,p_2)$ stands for the contraction of the
$\eps$-tensor with the 4-vectors in the argument.

\paragraph{\nlo{} virtual corrections}
The large-mass expansion~\cite{Smirnov:2002pj,Smirnov:1994tg},
formulated in terms of the ``method of regions'', states that the loop
integrand is to be Taylor-expanded in all relevant regions of loop
momenta, and the final result of each diagram is the sum over all those
expansions.  If a single internal mass $m$ is considered large, a loop
momentum $q^\mu_i$ can either be large, $|q^\mu_i|\sim m$, or
small, $|q^\mu_i|\ll m$, while external momenta $p^\mu_i$ are always
small, $|p^\mu_i|\ll m$.  In this way, the virtual \nlo{}
(two-loop) diagrams with top-quark loops reduce to either massive
two-loop vacuum integrals or products of massless one-loop triangles
with massive one-loop vacuum integrals.  As a result of the large-mass
expansion, all one-particle irreducible two-loop graphs involving
top-quark loops vanish, except for the ones with corrections to the
$\Pg\Pg \PG$ vertex.  Note that the latter as well as the axial vector
part of the $\Pg\Pg\PZ$ vertex each receive an anomalous
counterterm~\cite{Larin:1993tq} to restore chiral symmetry in the
massless-quark limit in the 't~Hooft--Veltman
scheme~\cite{'tHooft:1972fi,Breitenlohner:1977hr} for $\ga_5$, in the
following denoted as $\delta Z_5^{P}$ and $\delta Z_5^{A}$,
respectively.

Since we set $m_q=0$ for $q\ne\Pt$ both in the propagators and in the
Yukawa couplings, the set of diagrams to be evaluated with internal
massless quarks does not contain any two-loop box diagrams.  The only
genuine two-loop integrals with non-vanishing external momentum are
three-point functions as the one shown in \reffi{fig:diasb}\,(l). As 
their \lo{} counterparts at one loop, their contribution to the cross 
section vanishes when working in Landau gauge, where the
$\PZ$~propagator $D_{\mu\nu}(k)$ is proportional to the polarization sum
$\sum_\la$ for a physical vector particle $\PZ^\ast$ of mass $k^2$,
\begin{equation}
\begin{split}
D^{\mu\nu}(k) \propto -g^{\mu\nu}+\frac{k^\mu k^\nu}{k^2} = 
\sum_{\lambda}\varepsilon_{(\lambda)}^\mu(k)^*\varepsilon_{(\lambda)}^\nu(k)\,.
\end{split}
\end{equation}
The $\Pg\Pg\PZ^\ast{}$ subamplitude
therefore corresponds to the decay of a massive into two massless vector
particles which is forbidden due to the Landau--Yang
theorem~\cite{landau,Yang:1950rg}. 

In summary, the one-particle-irreducible two-loop diagrams comprise only
non-vanishing contributions from top-quark loops in the $\Pg\Pg \PG$ vertex
and massless quark loops in the $\Pg\Pg\PZ^*$ vertex. Since the contribution of
the latter vanishes in Landau gauge for the $\PZ$~boson, the genuine
two-loop calculation is particularly simple in that gauge.
It is, however, instructive to inspect the situation in general $R_\xi$ gauge
for the $\PZ$~boson as well, a task that is pursued in the Appendix.

The reducible two-loop graphs involve the product of the one-loop induced
$\Pg\Pg^*\PH$ and $\Pg\Pg^*\PZ$ vertices and
can be calculated
with conventional one-loop techniques. These graphs by themselves
form a gauge-invariant, UV-finite and IR-finite subset of diagrams.

In practice, we have performed two completely independent calculations
of all loop contributions.  Version~1 follows basically the strategy
described in \citere{Dawson:1998py} for the related process of
scalar--pseudoscalar Higgs-boson pair production. Here the Feynman
diagrams are generated with the program {\sc
  FeynArts}~\cite{Kublbeck:1990xc}, and the large-mass expansion of the
diagrams involving top-quark loops is performed by inhouse {\sc
  Mathematica} routines. The calculation is carried out in Landau
gauge, so that no two-loop diagrams with massless-quark loops
contribute.  In the second calculation, the diagrams are generated by
{\abbrev QGRAF}~\cite{Nogueira:1991ex} and expanded using {\abbrev
  EXP/Q2E}~\cite{Harlander:1997zb,Seidensticker:1999bb}. The massive
two-loop vacuum integrals resulting from top-quark loops are calculated
using {\abbrev MATAD}\cite{Steinhauser:2000ry}. This calculation is
carried out both in Landau and in unitarity gauge. In the latter, the 
Goldstone bosons are absent, but massless-quark loops
contribute to the $\Pg\Pg\PZ^*$ vertex; these graphs are calculated with
the program {\abbrev MINT}\cite{Harlander:2000mg}. 
While the results of the two
different calculations in Landau gauge are in mutual agreement term by term, 
for the calculation in unitarity gauge only the full result
agrees (after the renomalization procedure).

Including all required counterterms, the virtual contribution is given
by\footnote{Note that we consistently suppress terms involving $\ln
  4\pi$ and $\gamma_\text{E}$, which accompany poles in $\epsilon$,
  because they cancel in the UV-finite result as usual.}
\begin{equation}
\begin{split}
  \sigma_{virt} &= \int d\text{PS}_2 \, \left[
    \overline{|\M_0(\epsilon)|^2}( 1 + 2 \delta Z + \delta_\text{CS}) +
    2 \Re\left\{\overline{\M_1(\epsilon)\M_0(\epsilon)^\ast}\right\}
    \right]\,,
\label{eq:sigvirt}
\end{split}
\end{equation}
where $\M_0(\epsilon)$ is the \lo{} amplitude in $D=4-2\epsilon$
dimensions, $\M_1(\epsilon)$ the amplitude of the virtual corrections at
\nlo{}, and $d\text{PS}_2$ denotes the 2-particle phase-space
element. 
We renormalize the strong coupling in the $\msbar$ scheme
assuming $n_f=6$ flavours,
and the top-quark mass in the on-shell scheme. Note that
the renormalization factor is gauge dependent; in Landau gauge, it is
\begin{equation}
\begin{split}
  \delta Z & = Z_g^2 Z_3 Z_m^{-2\epsilon}Z_5^P -1 \\ &= 2\delta Z_g +
  \delta Z_3 -2\epsilon\delta Z_m + \delta Z_5^{P}, \\ \delta Z_g &=
  \api\Big(-\frac{11}{6}C_\text{A} + \frac{2}{3} T_\text{R} n_f\Big)
  \frac{1}{4\epsilon}, \\ \delta Z_3 &= -\api T_\text{R} \frac{1}{3}
  \bigg(\frac{1}{\epsilon} + \ln\Big(\frac{\mu^2}{\Mt^2}\Big)\bigg),
  \\ \delta Z_m &= -\api \Big(C_\text{F} \frac{3}{4\epsilon} + \mathcal
  O(1) \Big), \\ \delta Z_5^{P} &= -2\api C_\text{F},
\end{split}
\end{equation}
where $C_\text{F}=\frac{4}{3}$, $T_\text{R}=\frac{1}{2}$, and
$C_\text{A}=3$ are the \qcd{} colour factors. In unitarity gauge, the mass
counterterm $\delta Z_m$ is absent, but 
$\delta Z_5^{A}=-\api C_\text{F}$~\cite{Larin:1993tq}
is needed for the $\Pg\Pg\PZ$ vertex instead of $\delta Z_5^{P}$.

The term $\delta_\text{CS}$,
\begin{equation}
\begin{split}
  \delta_\text{CS} = \api \Big(\frac{\mu^2}{\hat s}\Big)^\epsilon
  &\bigg[C_\text{A} \Big(\frac{1}{\ep^2} - \frac{\pi^2}{3}\Big) +
    \Big(\frac{1}{\ep} + 1 \Big)\Big(\frac{11}{6}C_\text{A} -
    \frac{2}{3}T_\text{R} n_l\Big) \\ &+
    \Big(\frac{67}{18}-\frac{\pi^2}{6}\Big) C_\text{A}
    -\frac{10}{9}T_\text{R} n_l\bigg] \Big(1-\frac{\pi^2}{12}\ep^2\Big),
  \label{eq:deltaCS}
\end{split}
\end{equation}
in \refeq{eq:sigvirt} needs to be added
according to the dipole subtraction method \cite{Catani:1996vz},
where $n_l=5$ denotes the number of light flavours.

In order to be consistent with the currently available \pdf{} sets, we
express our final result in terms of $\alphas^{(5)}$, the strong
coupling with five active flavours using the matching relation
\cite{Chetyrkin:2000yt}
\begin{align}
  \alphas^{(5)}(\mu) &= \alphas^{(6)}(\mu) \Big[ 1 +
    \frac{\alphas^{(6)}(\mu)}{\pi}
    \Big(-\frac{1}{6}\ln\Big(\frac{\mu^2}{\Mt^2}\Big)\Big) + \mathcal
    O(\alphas^2) \Big],
\end{align}
so that the logarithmic dependence on $\Mt$ vanishes. 
This procedure is equivalent to decoupling
the top quark in the 
$\alphas$ renormalization by subtraction of the top-quark loop in the
gluon self energy at zero-momentum transfer, instead of using the
$\overline{\mathrm{MS}}$ prescription.
In the remainder
of this paper, we set $\alphas\equiv \alphas^{(5)}$.

Inserting the \qcd{} colour factors, the final result for the virtual
contribution can be written as
\begin{align}
 \sigma_{\mathrm{virt}} &= \bigg[ 1 + \frac{\alphas(\mu)}{\pi}
   \bigg( \frac{164}{9} + \frac{23}{6}\ln\Big(\frac{\mu^2}{\hat s}\Big)
   \bigg)\bigg] \sigma_\lo{} + \sigma_{(\mathrm{virt,red})},
\label{eq:sigma-virt}
\end{align}
where 
\begin{align}
  \sigma_\lo{} = \int d\text{PS}_2 \, \overline{|\M_0|^2},
\end{align}
(cf.\,\refeq{eq:M0square}) and $\sigma_{(\mathrm{virt,red})}$ denotes the
contribution from the type of reducible diagrams shown in \reffi{fig:diasb}\,(k).
It is not proportional to $\sigma_\lo{}$; as a function of the partonic
Mandelstam variables $\hat s=(p_1+p_2)^2$, $\hat t=(p_1-p_\PZ)^2$, and
$\hat u=(p_1-p_\PH)^2$, it reads
\begin{equation}
\begin{split}
&\sigma_{(\mathrm{virt,red})} = \int d\text{PS}_2 \, \Big(\api\Big)^3
\frac{\alpha^2\pi^2}{768\sw^4\cw^4\MZ^4}\frac{1}{\hat s-\MZ^2} 
\\ &\hspace{2em} \times \Bigg\{\big(\hat s\MZ^2\MH^2 +\hat s\MZ^4 -\hat
s^2\MH^2 -2 \hat s^2\MZ^2 +\hat s^3\big)  \\ &\hspace{3em}
\times \bigg(-2 + \ln\Big(\frac{-\hat u}{\MZ^2}\Big) \frac{\MZ^2}{\hat
  u-\MZ^2} + \ln\Big(\frac{-\hat t}{\MZ^2}\Big) \frac{\MZ^2}{\hat
  t-\MZ^2}\bigg)  \\ &\hspace{2em} + \big(-\hat s\MZ^2\MH^2 +
\hat s\MZ^4 + \hat s^2\MH^2 - 2\hat s^2\MZ^2 + \hat s^3 \big) 
\\ &\hspace{3em} \times\bigg(\frac{-\MZ^2}{\hat t-\MZ^2} +
\frac{-\MZ^2}{\hat u-\MZ^2} + \ln\Big(\frac{-\hat u}{\MZ^2}\Big)
\frac{\MZ^4}{(\hat u-\MZ^2)^2} + \ln\Big(\frac{-\hat t}{\MZ^2}\Big)
\frac{\MZ^4}{(\hat t-\MZ^2)^2} \bigg) \Bigg\}.
\end{split}
\end{equation}

\paragraph{\nlo{} real corrections}

The real corrections are induced by the partonic channels
$\Pg\Pg\to\PH\PZ\Pg$, 
$\Pg q\to\PH\PZ q$, 
$\Pg\bar q\to\PH\PZ\bar q$, and
$q\bar q\to\PH\PZ\Pg$, 
where in the channels involving external quarks only the squares of the
diagrams with closed quark loops are taken into account. At first sight,
the most complicated one-loop diagrams are pentagon graphs with a top-quark
loop. However, in the large-top-mass limit, these graphs vanish.
The algebraically most complicated diagrams are the box graphs with
external $\PZ^*\Pg\Pg\Pg$ fields; they
are the only ones that receive
contributions from the vector-coupling of the $\PZ$~boson, while 
all other diagrams (summed in pairs of opposite charge flow in the loop) 
depend only on the $\PZ$-boson axial-vector coupling. 

The actual calculation of the diagrams can be performed using standard
one-loop calculational techniques and has been carried out 
in three
completely independent ways.  The first approach
builds on graphs from {\sc FeynArts}~1.0~\cite{Kublbeck:1990xc} and
reduces or expands the full amplitudes with inhouse {\sc Mathematica}
routines, which produce output in the form of {\sc Fortran} code. The
occurring one-loop tensor and scalar integrals are numerically evaluated
with the (not yet public) library {\sc Collier} that is based on the
results of \citeres{'tHooft:1978xw,Passarino:1978jh,Beenakker:1988jr,
  Denner:2002ii,Denner:2005nn,Denner:2010tr}.  The second calculation is
based on the program packages {\sc FeynArts}~3.2~\cite{Hahn:2000kx} and
{\sc FormCalc}/{\sc LoopTools}~\cite{Hahn:1998yk,Hahn:2000jm}, as far as
the calculation of one-loop graphs is concerned that do not involve
top-quark loops. The large-mass expansion of the top-quark loops here
again is carried out using inhouse routines (independent from the ones
of version~1).

The third approach again uses the {\abbrev QGRAF/EXP/Q2E/MATAD}-setup
for the generation and expansion of the diagrams and the evaluation of
the massive vacuum diagrams. The calculation of the massless triangle
and box diagrams is performed by an extended version of the {\abbrev
  FORM}\cite{Vermaseren:2000nd} routine previously used in
\citere{Brein:2011vx}, which implements algebraic Passarino-Veltman
reduction\cite{Passarino:1978jh} and analytic results of the scalar
integrals given in~\citere{Ellis:2007qk}.  Here we explicitly verified
the gauge invariance of the result with respect to the $\PZ$~propagator
as well as to the external gluons.  For the latter, we assumed a general
axial gauge, where the polarization sum reads
  \begin{equation}
    \begin{split}
      \sum_{\lambda}\varepsilon_{(\lambda)}^{\mu,a}(p_i)^\ast
      \varepsilon_{(\lambda)}^{\nu,b}(p_i)
      = \delta^{ab}\left(-g^{\mu\nu} +\frac{p_i^\mu n^\nu + p_i^\nu
        n^\mu}{p_i\cdot n}\right) \quad (i=1,2,3)
    \end{split}
  \end{equation}
with an arbitrary light-like vector $n$ which drops out in the squared
amplitude.

All real-emission channels contain IR singularities in their integration
over phase space. More precisely, $\Pg\Pg\to\PH\PZ\Pg$ becomes IR singular
if the emitted gluon becomes soft or collinear to one of the incoming gluons.
The other channels involve only collinear singularities.
The separation of the IR singularities in the phase-space integration
is achieved using the standard
dipole subtraction method\cite{Catani:1996vz}, where an auxiliary cross section
is subtracted from the full real-emission part and added back after an 
analytical integration (in $D$ dimensions) over the one-particle emission phase that contains
the IR singularity (cf.\,\refeq{eq:deltaCS}).

\section{Numerical results}\label{sec:numerics}

\subsection{Input values}

We use the following input parameters:
\begin{equation}\begin{split}
& \MZ = 91.1876\GeV,\quad \MW = 80.399\GeV,\quad
G_\mu = 1.16637 \cdot 10^{-5}\GeV^{-2},
\\ 
& \mbottom =4.75\GeV,\qquad \mtop = 172\GeV.
\end{split}\end{equation}
In the effective-theory approximation, we set $\mtop\to\infty$
and $\mbottom=0$, as discussed above.  For the electromagnetic coupling
constant $\alpha$ we employ the $G_\mu$ scheme, where the coupling constant is
defined as
\begin{equation}
\begin{split}
\alpha = \frac{\sqrt{2}\,G_\mu \MW^2\sw^2}{\pi},\qquad
\sw^2 = 1-\cw^2=1-\frac{\MW^2}{\MZ^2}.
\end{split}
\end{equation}
As the default \pdf{} sets, we use {\abbrev MSTW}2008{\abbrev (N)LO}
\cite{Martin:2009iq} when evaluating a {\abbrev (N)LO} quantity. The
corresponding input values for the strong coupling are given by
$\alphas^\text{\lo}(\MZ) = 0.13939$ ($\alphas^\text{\nlo{}}(\MZ) =
0.12018$). The running of $\alphas$ is performed to the order under
consideration:
\begin{equation}
\begin{split}
\mu^2\deriv{\alphas}{\mu^2} = -\sum_{l=0}^n \alphas^{l+2}\beta_l\,,
\end{split}
\end{equation}
with $n=0$ ($n=1$) at \lo{} (\nlo{}). Both the \pdf{}s and the
$\alphas$ evolution are implemented with the help of the {\abbrev
  LHAPDF} library\cite{lhapdf}.

Our default choice for the renormalization and the factorization scales
$\muR$ and $\muF$ is the invariant mass of the $\PH\PZ$ system:
\begin{equation}
\begin{split}
\mu_0 = \sqrt{(p_\PH + p_\PZ)^2}\,.
\label{eq:centralmu}
\end{split}
\end{equation}

\subsection{Leading-order considerations}\label{sec::lo}

\begin{figure}
\begin{center}
\includegraphics[width=\textwidth]{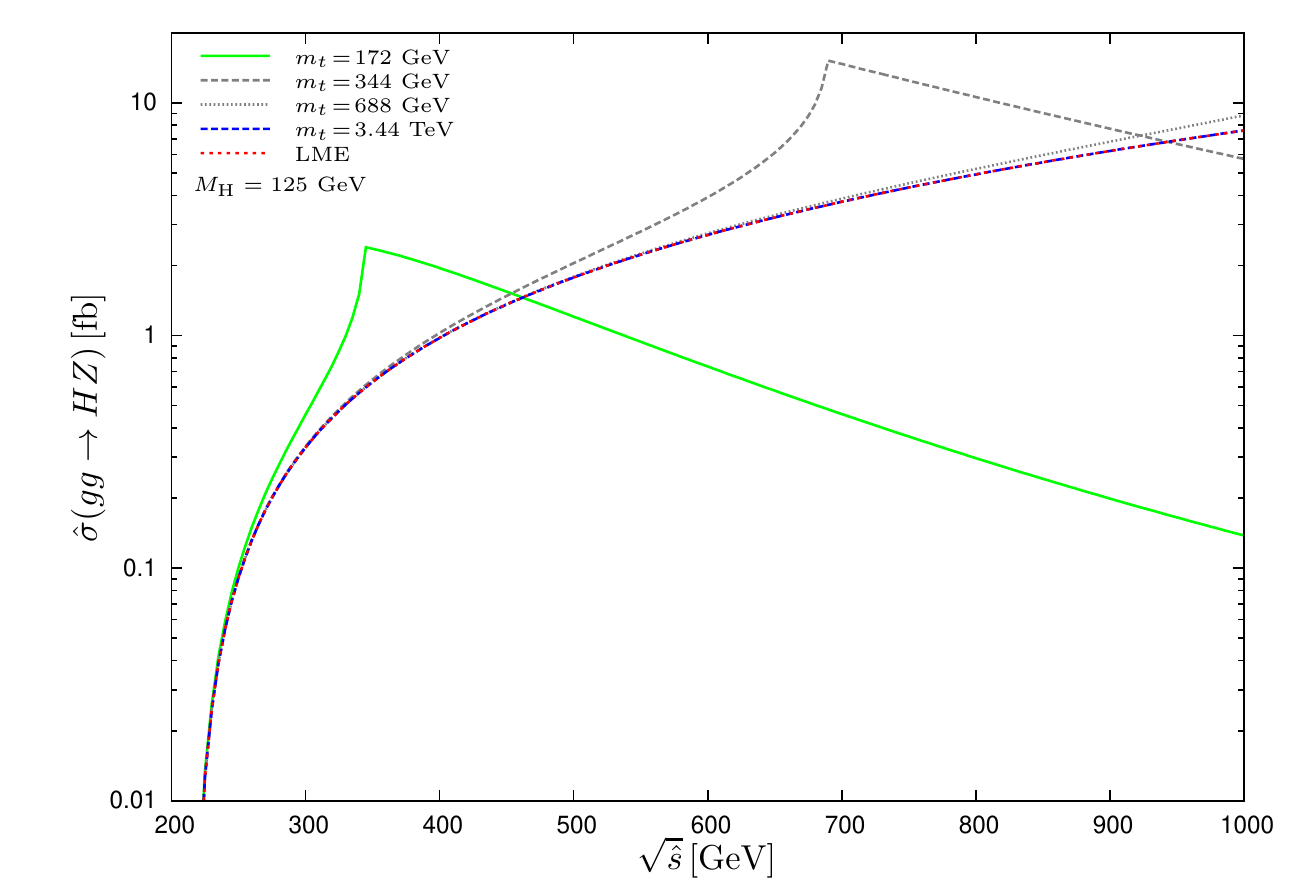}
\caption[]{\label{fig:loparton} Comparison of the \lo{} partonic cross
  section in the effective (labelled ``{\abbrev LME}'') and the full
  theory for various, partly hypothetical
  values of the top-quark
  mass. The curve for $\mtop=3.44$\,TeV cannot be distinguished from the
  {\abbrev LME} result.}
\end{center}
\end{figure}

In order to get an idea about the quality of the effective theory, we
show some studies at \lo{} before presenting our \nlo{}
results. Figure\,\ref{fig:loparton}
shows the partonic cross section both for
the exact top-mass dependence and in the effective theory. The exact
result exhibits a kink at the top-quark pair threshold $\sqrt{\hat s} =
2\mtop = 344\GeV$ which clearly cannot be reproduced by the 
effective-theory approach. For larger values of $\sqrt{\hat s}$, we do not expect
an expansion in $1/\mtop$ to converge. In fact, higher-order terms in
this expansion would most likely worsen the prediction in the region of larger
$\sqrt{\hat s}$.

\begin{figure}
\begin{center}
  \subfigure[Inclusive cross section]{\includegraphics[width=0.49\textwidth]{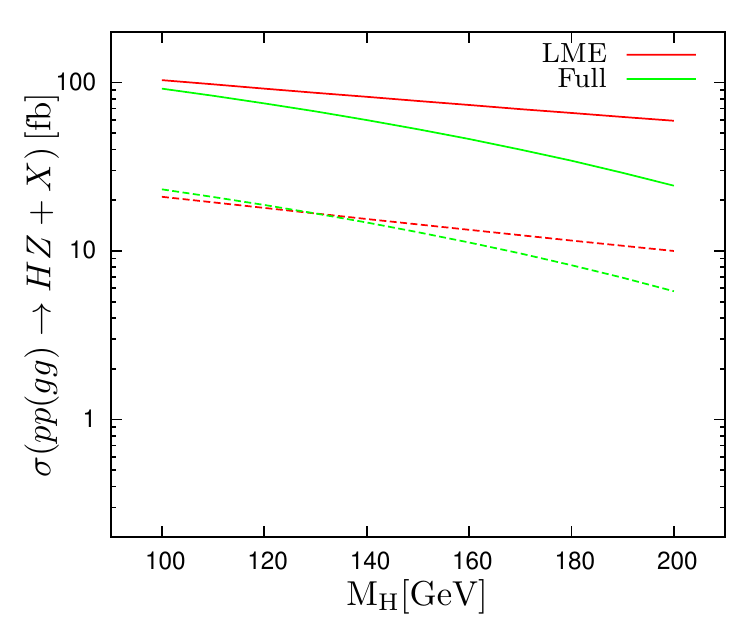}}
  \subfigure[$\ptH{}>200\GeV$]{\includegraphics[width=0.49\textwidth]{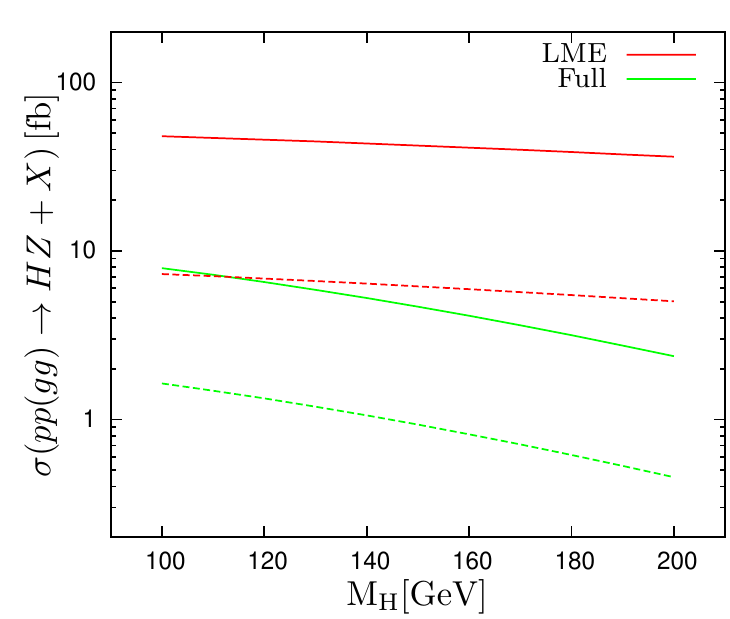}}
  \caption[]{\label{fig:lohadron}%
    Comparison of the \lo{} hadronic cross section in the effective and the full 
    theory for $\sqrt{s}=8\TeV$ (dashed) and $14\TeV$ (solid).}
\end{center}
\end{figure}

Taking into account the kinematical constraint $\sqrt{\hat s}>
\mhiggs+\MZ$, the region where the effective theory is nominally
applicable shrinks to zero for $\mhiggs>2\mtop-\MZ\approx 253\GeV$.
Figure\,\ref{fig:lohadron}\,(a) compares the total inclusive \lo{}
hadronic
cross section at $8\TeV$ and $14\TeV$ when the full top- and
bottom-mass dependence is taken into account to the effective-theory
result. The behaviour is expected from the considerations above: The
effective theory works better for smaller Higgs masses, agreeing to the
full results within 2\% (25\%) for $8\TeV$ ($14\TeV$) at
$\mhiggs=125\GeV$.  Note that the \pdf{}s suppress the contribution from
larger $\hat s$, thus emphasising the region where the $1/\mtop$
expansion converges.  For larger values of $\mhiggs$, the
effective-theory approximation deteriorates; at $\mhiggs=200\GeV$, the
deviation to the full result is 74\% (143\%) for $8\TeV$ ($14\TeV$).

The situation becomes more problematic in the boosted regime which we
study by imposing a lower cut on the Higgs' transverse momentum,
requiring $\ptH{}>200\GeV$, see \reffi{fig:lohadron}\,(b).
 In this case,
the minimal value for $\sqrt{\hat s}$ is already above the top-quark
threshold when $\mhiggs=100\GeV$. Consequently, the direct application
of the effective-theory approximation is off by almost a factor of five
to ten, which is clearly unacceptable.

A direct evaluation of the \nlo{} contribution in the effective theory
is therefore not possible. However, in
\citeres{Spira:1995rr,Kramer:1996iq,Harlander:2010my,Harlander:2009bw,%
  Pak:2009dg,Pak:2011hs,Harlander:2012hf} it was shown for the process
$\Pg\Pg\to \PH$ at \nlo{} and \nnlo{} that the perturbative {\it
  correction factor}, defined at \nlo{} in \refeq{eq:kfacdef}, depends
only very weakly on the top-quark mass. To some degree, this holds even
far outside the convergence region of the heavy-top expansion, as long
as only the leading term in $1/\mtop$ is taken into account. Motivated
by this observation, we move on to \nlo{} and present our results in the
next section.

\subsection{Next-to-leading order results}\label{sec::nlo}

\subsubsection{Correction factor}\label{sec:corrfac}

As outlined above, we evaluate the \nlo{} hadronic cross section by
rescaling the full \lo{} result by the perturbative $K$-factor
calculated in the effective theory:
\begin{equation}
\begin{split}
\sigma^\nlo{}_{\mathrm{approx}}(\mtop,\mbottom) &=
\sigma^\lo{}(\mtop,\mbottom)
K(\mtop\to\infty,\mbottom=0)\\
&= \frac{\sigma^\lo{}(\mtop,\mbottom)}
{\sigma^\lo{}(\mtop\to\infty,\mbottom=0)}\,
\sigma^\nlo{}(\mtop\to\infty,\mbottom=0)\,.
\label{eq:sigmanlo}
\end{split}
\end{equation}
Since we are aiming at a \nlo{} quantity, it actually might be more appropriate
to evaluate the formally \lo{} cross sections in \refeq{eq:sigmanlo} with
\nlo{} \pdf{}s. We checked that the effect of this is much smaller than
the uncertainty due to variations of the renormalization and
factorization scale, which is why we stick to \lo{} \pdf{}s in $\sigma^\lo{}$.

\begin{figure}
\begin{center}
  \subfigure[Inclusive cross
    section]{\includegraphics[width=0.49\textwidth]{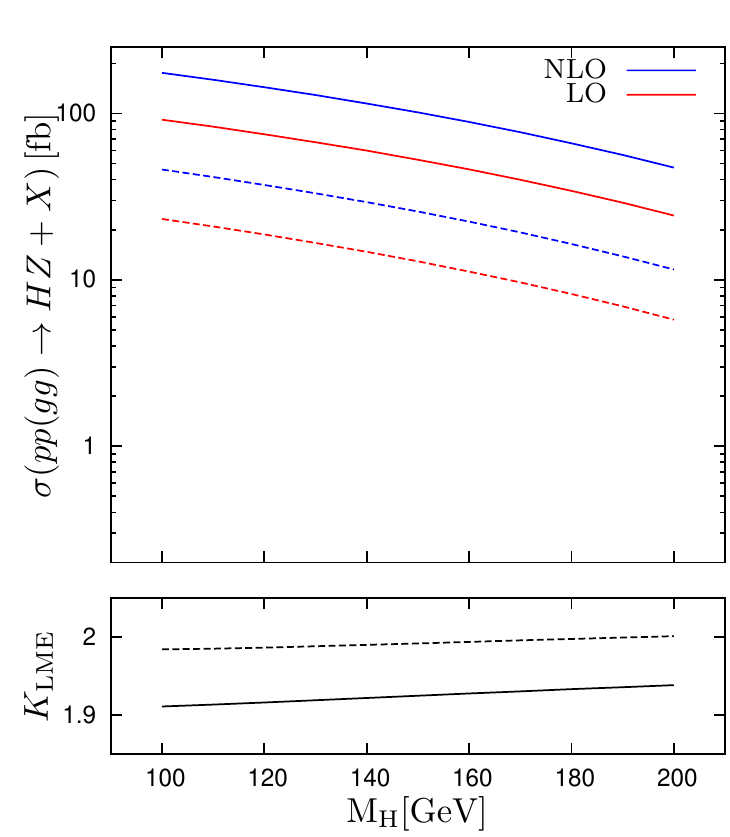}}
  \subfigure[$\ptH{}>200\GeV$]{\includegraphics[width=0.49\textwidth]{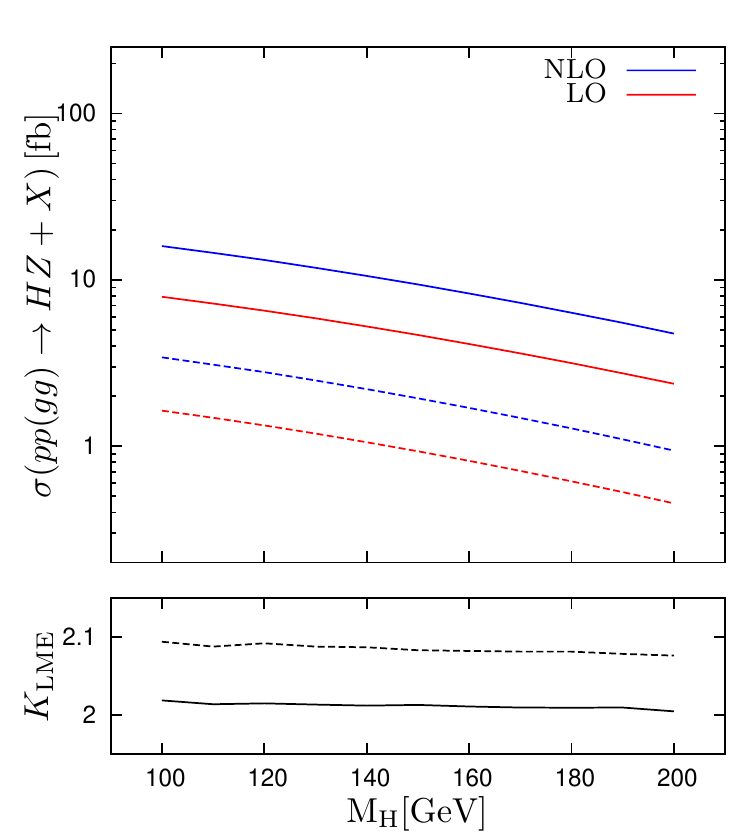}}
  \caption[]{\label{fig:nlotot} \nlo{} hadronic cross section as
    obtained by using \refeq{eq:sigmanlo} (upper), and \nlo{} $K$-factor
    (lower) for $\sqrt{s}=8\TeV$ (dashed) and $14\TeV$ (solid).}
\end{center}
\end{figure}
Figure~\ref{fig:nlotot}
shows the gluon-induced cross section obtained
in this way for $\sqrt{s}=8\TeV$ and $\sqrt{s}=14\TeV$ hadronic
centre-of-mass energy, together with the corresponding perturbative
correction factor $K$. Part (a) of \reffi{fig:nlotot} shows the total
inclusive cross section, while in part (b) the boosted scenario with
$\ptH{}>200\GeV$ is shown. In both cases, we observe a $K$-factor of the
order of two, almost independent of $\mhiggs$, with a slight increase towards 
lower centre-of-mass energies. This behaviour is very similar
to the one observed for gluon-induced single-Higgs%
\cite{Spira:1995rr,Kramer:1996iq,Harlander:2010my,Harlander:2009bw,%
  Pak:2009dg,Pak:2011hs,Harlander:2012hf} and Higgs pair
production~\cite{Dawson:1998py}.  The correction even slightly exceeds
the well-known correction factor for $\Pg\Pg\to \PH$.  

\begin{figure}
\begin{center}
  \includegraphics[width=\textwidth]{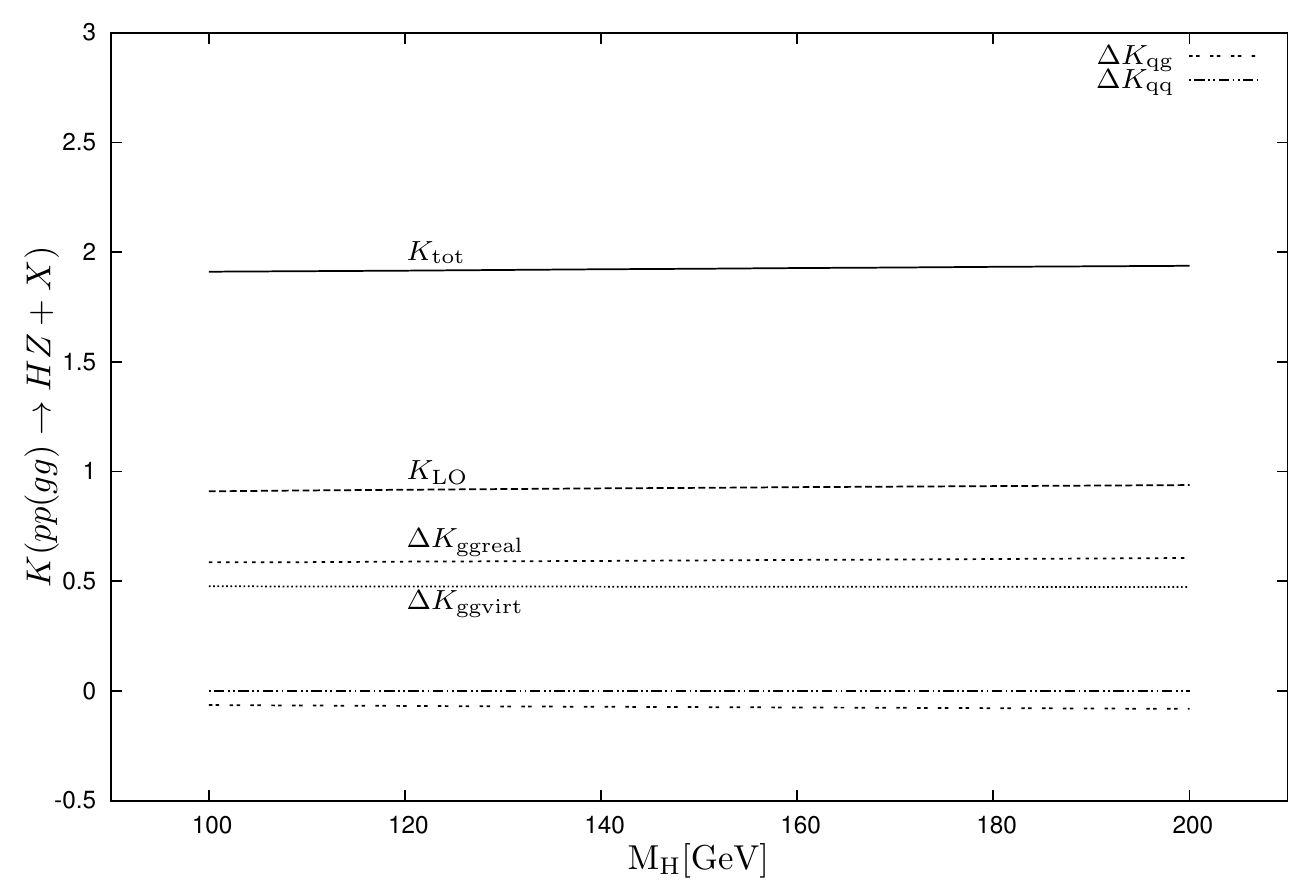}
  \caption[]{\label{fig:ksplit} Individual contributions to the \nlo{}
    hadronic $K$-factor at $\sqrt{s}=14$\,TeV as discribed in the main
    text.}
\end{center}
\end{figure}

A breakdown into individual contributions to $K$ is shown in \reffi{fig:ksplit} for the
total inclusive cross section at $\sqrt{s} = 14\TeV$:
\begin{itemize}
\item $K_\text{\lo}$ --- change of \pdf{} sets from \lo{} to \nlo{}
\item $\Delta K_\text{\text{ggvirt}}$ --- 
 virtual corrections including integrated
  dipole terms according to \citere{Catani:1996vz}
\item $\Delta K_{\text{ggreal}}$ --- correspondingly regularized real corrections
\item $\Delta K_\text{qg}, \Delta K_\text{qq}$ --- contributions from
  $qg$ and $q\bar q$ initial states
\end{itemize}
The sum of all these terms results in $K_\text{tot}$, the total $K$-factor.

\subsubsection{Residual scale uncertainty}\label{sec:resscale}

\begin{figure}
\begin{center}
  \subfigure[Inclusive cross section]{\includegraphics[width=0.49\textwidth]{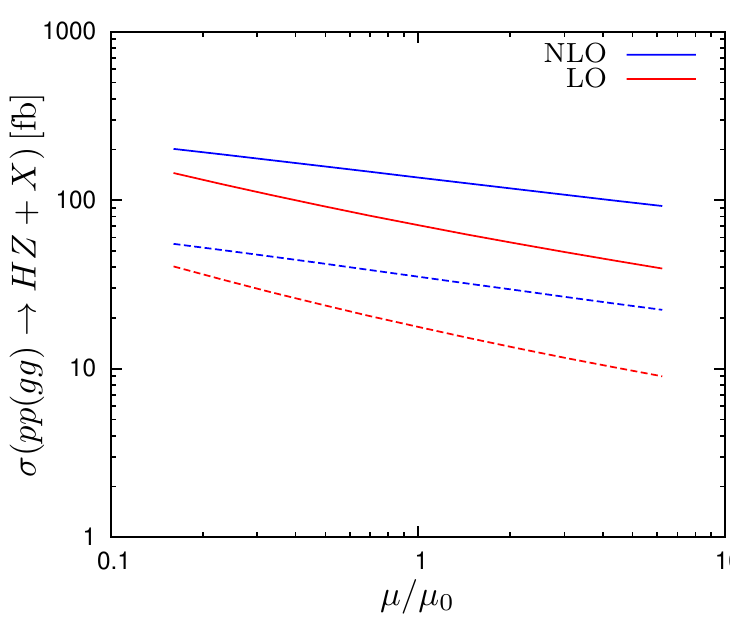}}
  \subfigure[$\ptH{}>200\GeV$]{\includegraphics[width=0.49\textwidth]{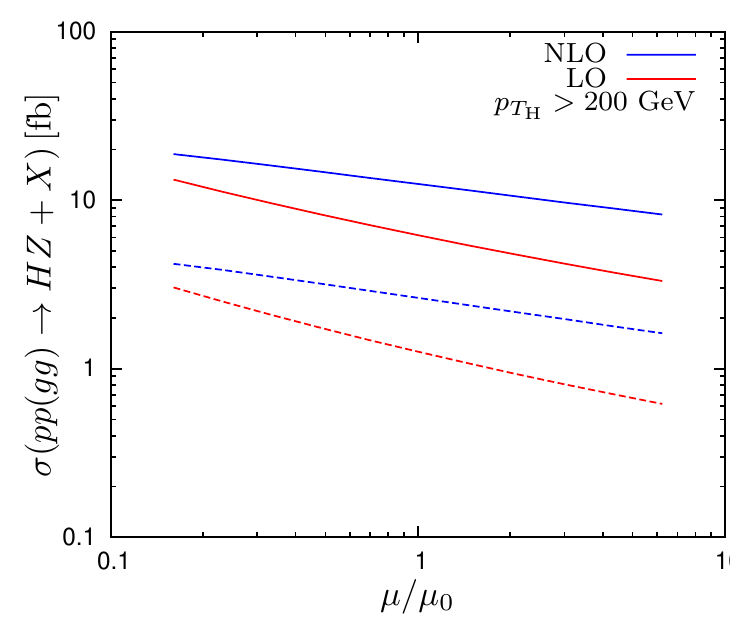}}
   \caption[]{\label{fig:scales} Scale dependence of the hadronic \lo{}
     and \nlo{} cross section for $\sqrt{s}=8\TeV$ (dashed) and $14\TeV$
     (solid).  The renormalization and factorization scales are varied
     simutaneously around the central scale
     $\mu_0=\sqrt{(p_\PH+p_\PZ)^2}$. The Higgs mass is set to
     $\MH=125$\,GeV.}
\end{center}
\end{figure}

As described in the introduction, the \lo{} scale dependence for this
purely gluon-induced process is quite large. \nlo{} corrections
typically decrease this uncertainty. Let us recall the situation in the
gluon-fusion process $\Pg\Pg\to \PH$, however: For the \lo{} result, the
usually adopted scale variation by a factor of two around the central
scale leads to a gross underestimation of the size of the higher-order
effects. At \nlo{}, the scale uncertainty is not significantly smaller
than at \lo{}, but it does provide a good estimate of the \nnlo{}
effects. Consistently, inclusion of the \nnlo{} corrections leads
to a significant reduction of the scale uncertainty.

Expecting a similar behaviour for the $\Pg\Pg\to \PH\PZ$ process, it is
not surprising to see the result shown in \reffi{fig:scales}: Both for the
inclusive and the boosted scenario the scale dependence decreases from
more than 100\% at \lo{} to 60\% at \nlo{} when the renormalization and
factorization scales are varied simultaneously by a factor of six around
their central value $\mu_0$, see \refeq{eq:centralmu}. As for the process
$\Pg\Pg\to \PH$, the behaviour in $\mu/\mu_0$ is strictly monotonous,
and the \lo{} and \nlo{} curves do not intersect. Therefore,
a preferred value for
$\muF$ and $\muR$ cannot be deduced from these plots. The radiative
corrections increase with $\mu/\mu_0$, so there is a slight tendency
towards choosing smaller values of $\mu$. Nevertheless, in our numerical
analysis we stick to the ``natural'' value $\mu_0$ as the central
choice.

Note that, also similar to what is observed in $\Pg\Pg\to \PH$, variation by a
factor of two would not lead to any overlap between the \lo{} and the
\nlo{} predictions.

The similarity between the processes $\Pg\Pg\to \PH$ and $\Pg\Pg\to
\PH\PZ$ suggests that the \nlo{} error estimate due to scale variation
is quite reliable for the process $\Pg\Pg\to \PH\PZ$. In order to take
into account the fact that the effective theory is expected to work not
quite as well in $\Pg\Pg\to \PH\PZ$ as in $\Pg\Pg\to \PH$, we determine
this uncertainty by varying $\mu$ within a factor of three rather than
two around the central value $\mu_0$. The numerical results are listed
in \refta{tab:numberscrosssect}.

\begin{table}
  \centering
  \begin{tabular}{|cc|c|c|}\hline
  $\sqrt{s}$~[TeV]& $\MH$[GeV] &$\phantom{abcde}$ $\sigma_{\Pg\Pg}^{\lo{}}$[fb]
    $\phantom{abcde}$&$\phantom{abcde}$$\sigma_{\Pg\Pg}^{\nlo{}}$[fb]
    $\phantom{abcde}$\\\hline
   \multicolumn{2}{|c}{no $\ptH{}$ cut}&\multicolumn{2}{c|}{}\\\hline
 $8$  &  $115$  &  $  19.8^{+61\%}_{-34\%}$ &  $
39.3^{+32\%}_{-24\%}$\\[4pt]
 $8$  &  $120$  &  $  18.7^{+61\%}_{-34\%}$ &  $
37.2^{+32\%}_{-24\%}$\\[4pt]
 $8$  &  $125$  &  $  17.7^{+61\%}_{-34\%}$ &  $
35.1^{+32\%}_{-24\%}$\\[4pt]
 $8$  &  $130$  &  $  16.7^{+61\%}_{-34\%}$ &  $
33.1^{+32\%}_{-24\%}$\\[4pt]
 $14$  &  $115$  &  $  79.1^{+51\%}_{-31\%}$ &  $
152^{+27\%}_{-21\%}$\\[4pt]
 $14$  &  $120$  &  $  75.1^{+51\%}_{-31\%}$ &  $
144^{+27\%}_{-21\%}$\\[4pt]
 $14$  &  $125$  &  $  71.1^{+51\%}_{-31\%}$ &  $
136^{+27\%}_{-21\%}$\\[4pt]
 $14$  &  $130$  &  $  67.2^{+51\%}_{-31\%}$ &  $
129^{+27\%}_{-21\%}$\\[4pt]

\hline
  \multicolumn{2}{|c}{$\ptH{}>200$~GeV}&\multicolumn{2}{c|}{}\\\hline
 $8$  &  $115$  &  $  1.41^{+65\%}_{-36\%}$ &  $
 2.94^{+34\%}_{-25\%}$\\[4pt]
 $8$  &  $120$  &  $  1.33^{+65\%}_{-36\%}$ &  $
 2.79^{+33\%}_{-26\%}$\\[4pt]
 $8$  &  $125$  &  $  1.26^{+65\%}_{-36\%}$ &  $
 2.63^{+34\%}_{-25\%}$\\[4pt]
 $8$  &  $130$  &  $  1.19^{+65\%}_{-36\%}$ &  $
 2.48^{+33\%}_{-25\%}$\\[4pt]
 $14$  &  $115$  &  $  6.86^{+55\%}_{-32\%}$ &  $
 13.8^{+29\%}_{-22\%}$\\[4pt]
 $14$  &  $120$  &  $  6.53^{+55\%}_{-32\%}$ &  $
 13.1^{+28\%}_{-22\%}$\\[4pt]
 $14$  &  $125$  &  $  6.19^{+55\%}_{-32\%}$ &  $
 12.5^{+29\%}_{-22\%}$\\[4pt]
 $14$  &  $130$  &  $  5.87^{+55\%}_{-32\%}$ &  $
 11.8^{+29\%}_{-22\%}$\\[4pt]

\hline
  \end{tabular}
  \caption{Cross sections of $\PH\PZ$ production via gluon fusion for
    \lhc{} energies in the range of phenomenologically preferred $\MH$
    values. The scale uncertainty is given in percent.  The latter
    results from a rescaling of $\muR=\muF$ by factors of 3 and 1/3
    relative to $\mu_0$.}
  \label{tab:numberscrosssect}
\end{table}

\subsubsection{Total inclusive cross section}

In this section we provide the most up-to-date numbers for the
total inclusive cross section for the Higgs-strahlung process at the
\lhc{} with $8$ and $14\TeV$, including
\begin{itemize}
\item \nnlo{} Drell--Yan terms $\sigma^\text{HV,DY}$
 of order
  $g^4\alphas^n$
  ($n=0,1,2$)~\cite{Hamberg:1991np,Harlander:2002wh,Brein:2003wg};
\item electroweak corrections which are applied as an overall factor to
  the Drell--Yan terms\cite{Ciccolini:2003jy,Brein:2004ue};
\item top-loop-induced corrections of order
  $\order{\lambda_{\Pt}g^3\alphas^2}$~\cite{Brein:2011vx};
\item gluon-induced terms of order $\lambda_{\Pt}^2g^2\alphas^n$
  ($n=2,3$); $n=3$ corresponds to the newly calculated terms of this paper.
\end{itemize}
For the non-gluon-fusion part of the cross section, the scale variation
is obtained by using the {\abbrev MSTW}2008\nnlo{} \pdf{} set and
varying $\muF$ and $\muR$ independently within the interval
$(\muR,\muF)/\mhiggs\in [1/3,3]\times[1/3,3]$, which results in a cross
section interval $[\sigma_{\text{no-}\Pg\Pg}^{(-)},
  \sigma_{\text{no-}\Pg\Pg}^{(+)}]$.  The central value of the total
cross section is then obtained as
\begin{equation}
\begin{split}
\sigma_\text{central} &=
\frac{1}{2}\left[\sigma_{\text{no-}\Pg\Pg}^{(+)}
+\sigma_{\Pg\Pg}^{(+)}
+\sigma_{\text{no-}\Pg\Pg}^{(-)}
+\sigma_{\Pg\Pg}^{(-)}
\right],
\end{split}
\end{equation}
where $\sigma_{\Pg\Pg}^{(\pm)}$ are the boundaries of the
scale uncertainty interval of the gluon-induced component which can be
obtained from \refta{tab:numberscrosssect}.  Accordingly, the scale
uncertainty is calculated as
\begin{equation}
\begin{split}
\Delta_\text{scale} &=
\left[\sigma_{\text{no-}\Pg\Pg}^{(+)}
+\sigma_{\Pg\Pg}^{(+)}
-\sigma_{\text{no-}\Pg\Pg}^{(-)}
-\sigma_{\Pg\Pg}^{(-)}
\right]/(2\sigma_\text{central})\,.
\end{split}
\end{equation}

The influence of the newly evaluated \nlo{} gluon-induced terms on the
overall \pdf{}$+\alphas$ uncertainty is rather small, since this
contribution comprises only about 5\% of the total cross section, and
will be neglected.  Therefore, we base the estimate of the
\pdf{}$+\alphas$ uncertainty solely on what is currently contained in
    {\tt vh@nnlo} (i.e., \lo{} gluon-induced terms {\it are} taken into
    account).  Following the {\abbrev PDF4LHC}
    recommendations\,\cite{Botje:2011sn} by using the \nnlo{} \pdf{}
    sets from {\abbrev MSTW}2008\,\cite{Martin:2009iq}, {\abbrev
      CT}10\,\cite{Lai:2010vv}, and {\abbrev
      NNPDF}23\,\cite{Ball:2012cx}, we obtain a cross section interval
    $[\sigma_{\text{\pdf}+\alphas}^{(-)},\sigma_{\text{\pdf}+\alphas}^{(+)}]$
    and calculate the resulting uncertainty as
\begin{equation}
\begin{split}
\Delta_{\text{\pdf}+\alphas} &=
\frac{\sigma_{\text{\pdf}+\alphas}^{(+)}-\sigma_{\text{\pdf}+\alphas}^{(-)}}
     {\sigma_{\text{\pdf}+\alphas}^{(+)}+\sigma_{\text{\pdf}+\alphas}^{(-)}}\,.
\end{split}
\end{equation}

\begin{table}
\begin{center}
\begin{tabular}{|cc|r@{.}l|c|c|c|}
\hline
$\sqrt{s}$\,[TeV]& $\MH$\,[GeV] &\multicolumn{2}{c|}{$\sigma$[pb]}
    & $\Delta_{\text{scale}}$[\%] & $\Delta_{\text{\pdf}+\alphas}$[\%] &
$\Delta_\text{total}$[\%] \\\hline
\multicolumn{7}{|c|}{$\Pp\Pp\to\PH\PW$}\\
\hline
 $8$ & $115$ &  0&926 & $\pm  0.6$ & $\pm  2.3$ & $\pm  2.9$ \\
 $8$ & $120$ &  0&805 & $\pm  0.6$ & $\pm  2.5$ & $\pm  3.1$ \\
 $8$ & $125$ &  0&705 & $\pm  0.6$ & $\pm  2.3$ & $\pm  3.0$ \\
 $8$ & $130$ &  0&617 & $\pm  0.7$ & $\pm  2.4$ & $\pm  3.1$ \\
 $14$ & $115$ &  1&97 & $\pm  0.6$ & $\pm  2.0$ & $\pm  2.6$ \\
 $14$ & $120$ &  1&73 & $\pm  0.7$ & $\pm  1.8$ & $\pm  2.5$ \\
 $14$ & $125$ &  1&52 & $\pm  0.7$ & $\pm  2.2$ & $\pm  2.9$ \\
 $14$ & $130$ &  1&34 & $\pm  0.6$ & $\pm  2.0$ & $\pm  2.6$ \\

\hline
\multicolumn{7}{|c|}{$\Pp\Pp\to\PH\PZ$}\\
\hline
 $8$ & $115$ &  0&540 & $\pm  2.6$ & $\pm  2.4$ & $\pm  5.0$ \\
 $8$ & $120$ &  0&475 & $\pm  2.8$ & $\pm  2.4$ & $\pm  5.1$ \\
 $8$ & $125$ &  0&419 & $\pm  2.9$ & $\pm  2.4$ & $\pm  5.3$ \\
 $8$ & $130$ &  0&371 & $\pm  3.1$ & $\pm  2.3$ & $\pm  5.4$ \\
 $14$ & $115$ &  1&24 & $\pm  3.5$ & $\pm  1.8$ & $\pm  5.3$ \\
 $14$ & $120$ &  1&10 & $\pm  3.7$ & $\pm  1.6$ & $\pm  5.3$ \\
 $14$ & $125$ &  0&983 & $\pm  3.9$ & $\pm  1.6$ & $\pm  5.4$ \\
 $14$ & $130$ &  0&880 & $\pm  4.0$ & $\pm  1.9$ & $\pm  5.9$ \\

\hline
\end{tabular}
\caption[]{\label{tab:xsec}Total inclusive cross section for the processes
  $\Pp\Pp\to \PH\PW$ and $\Pp\Pp\to \PH\PZ$. The latter includes the
  newly calculated \nlo{} gluon-induced terms. The evaluation of the
  scale and \pdf{}+$\alphas$ uncertainties is described in the main text.}
\end{center}
\end{table}
Our results are shown in \refta{tab:xsec}.  We find that the \nlo{}
gluon-induced terms calculated in this paper increase the central values
of the $\PH\PZ$ cross section by about $4\%$ ($7\%$) at 8\,TeV
(14\,TeV). Since the $K$-factor for these terms is of the order of two,
and their scale uncertainty decreases by roughly the same factor when
going from \lo{} to \nlo{} (see \refta{tab:numberscrosssect}), the
overall scale uncertainty on the total inclusive cross section remains
almost unaffected by the inclusion of the new terms. For completeness,
we also include updated numbers for $\PH\PW$ production in
\refta{tab:xsec}, even though they are not affected by the \nlo{}
gluon-fusion terms calculated in this paper.

As a side remark, we note that the \pdf{}$+\alphas$ uncertainties of
\refta{tab:xsec} are significantly smaller than in
\citere{Brein:2011vx}. This is due to the use of only \nnlo{} \pdf{}
sets in this newer version, while the previous numbers were based on a
rescaling of the \nnlo{} {\abbrev MSTW}2008 uncertainty by the \nlo{}
\pdf{} error. For $\PH\PW$ production, also the scale uncertainty is
slightly smaller in \refta{tab:xsec} than in \citere{Brein:2011vx}. This
is because these previous numbers were obtained by linearly adding
uncertainties of the ``top-induced'' terms to the rest, while here we
vary the scale in both contributions simultaneously. For the $\PH\PZ$
cross section, this is overcompensated by the uncertainty of the
\nlo{} gluon-fusion component, see above.

\section{Conclusions}\label{sec::conclusions}

The gluon-induced corrections to the Higgs-strahlung process have been
calculated through \nlo{}, i.e.\ $\order{\alphas^3}$. The perturbative
correction factor is obtained in the limit $\mtop\to \infty$ and
$\mbottom=0$, and then used to rescale the full \lo{} cross section in
order to obtain a prediction for the \nlo{} result. The behaviour of the
perturbative series and the residual scale variation are found to be
comparable to the gluon-fusion processes of single-Higgs and Higgs pair
production. The success of the effective-field-theory approach in
describing corrections to single-Higgs productions, where
exact calculations for higher-order corrections are available,
gives confidence in a reliable prediction
of the theoretical uncertainty due to higher-order effects
for gluon-induced $\PH\PZ$ production.

Numerical results were provided for current and future \lhc{} energies,
both for the fully inclusive cross section and for boosted Higgs
kinematics.  We use these results in order to provide the most
up-to-date predictions for the hadronic Higgs-strahlung process at
relevant collider energies. The large corrections on the gluon-induced
terms, combined with their large scale uncertainty, increases the
overall uncertainty on the total $\PH\PZ$ cross section by about 1\%.

In the near future, the results will be included in the publically
available numerical program {\tt vh@nnlo}.

\section*{Appendix}

\section*{\boldmath{Treating $\Pg\Pg\to\PH\PZ$ via the anomaly relation}}

In this appendix we sketch yet another way to calculate the genuine
two-loop part upon employing various tricks, such as the Landau--Yang
theorem~\cite{landau,Yang:1950rg}, the Adler--Bell--Jackiw ({\abbrev
 ABJ}) anomaly relation~\cite{Adler:1969gk}, and the Adler--Bardeen
theorem~\cite{Adler:1969er}, in order to reduce the calculation to a
much simpler massless one-loop calculation.  For definiteness we employ
the conventions of \citere{Denner:1991kt} for the \sm{} parameters and
Feynman rules.

In \refse{sec:details} we explained that the \lo{} and \nlo{} virtual
amplitudes for the process $\Pg\Pg\to\PH\PZ$ have the following properties:
\begin{itemize}
\item
Non-vanishing contributions originate only 
from loops of vertex types $\Pg\Pg\PZ$, $\Pg\Pg \PG$, and $\Pg\Pg\PH$,
as illustrated in \reffi{fig:diasb}(b,c,k,l,m),
while all graphs with more than three 
external legs attached to the loop vanish or compensate each other
for $\Mt\to\infty$ and $m_{\Pq\ne\Pt}=0$.
\item
At \nlo{} the loop-induced $\Pg\Pg\PH$ vertex is only relevant at the one-loop
level within reducible diagrams like the one shown in \reffi{fig:diasb}(k).
Owing to their simplicity we do not consider those reducible diagrams in the
following.
\item
When the $\Pg\Pg\PZ$ vertex is attached to an intermediate
$\PZ$~propagator, and only in this case two-loop diagrams of this vertex
become relevant, the Landau--Yang theorem implies that only the
longitudinal part of the $\PZ$~propagator contributes.
\end{itemize}
Thus, the contribution of one-particle-irreducible (1PI) diagrams, i.e.\ the graphs in 
\reffi{fig:diasb}\ (b,c,l,m), to the matrix elements $\M_n$ ($n=0,1$) is given by
\begin{equation}\begin{split}
\M_n^{\mathrm{1PI}}(\eps) &= \veps_1^\mu\veps_2^\nu \, G_{\mu\nu\rho}^{\Pg^a\Pg^b\PZ}(p_1,p_2,-k)
\, D^{\rho\si}_\xi(k) \, \frac{e\MZ}{\sw\cw} \, \veps_{\PZ,\si}^*
\\
           &  -\ri\veps_1^\mu\veps_2^\nu \, G_{\mu\nu}^{\Pg^a\Pg^b\PG}(p_1,p_2,-k)
\, D_\xi(k) \, \frac{e}{2\sw\cw} \, (p_\PZ+2p_\PH)^\si \,\veps_{\PZ,\si}^*,
\label{eq:M1PI}
\end{split}\end{equation} 
where we use an obvious notation for the external polarization
vectors $\veps_1, \veps_2, \veps_{\PZ}^*$ for the two incoming gluons
and the outgoing Z~boson, $a,b$ are the gluonic colour indices,
$k=p_1+p_2$ is the momentum of the intermediate $\PZ$ or $\PG$ lines,
$G_{\mu\nu\rho}^{\Pg^a\Pg^b\PZ}(p_1,p_2,-k)$ and
$G_{\mu\nu}^{\Pg^a\Pg^b\PG}(p_1,p_2,-k)$ are the amputated Green
functions for the $\Pg\Pg\PZ/\PG$ vertices, and $D^{\rho\si}_\xi(k)$ and
$D_\xi(k)$ are the $\PZ$/$\PG$ propagators in the general $R_\xi$ gauge,
\beq D^{\rho\si}_\xi(k) = \frac{-\ri\left(g^{\rho\si}-\frac{k^\rho
    k^\si}{k^2}\right)}{k^2-\MZ^2} -\ri\frac{k^\rho
  k^\si}{k^2}\frac{\xi}{k^2-\xi\MZ^2}, \qquad D_\xi (k) =
\frac{\ri}{k^2-\xi\MZ^2}.
\label{eq:props}
\eeq
Inserting the latter into \refeq{eq:M1PI} and exploiting the fact that the transversal part of 
$D^{\rho\si}_\xi(k)$ (first term in \refeq{eq:props}) does not contribute,
we obtain
\begin{equation}
\begin{split}
\M_n^{\mathrm{1PI}}(\eps) &= -\frac{e}{\sw\cw}
\,\frac{(p_\PH\cdot\veps_{\PZ}^*)}{\hat s-\xi\MZ^2} \, \veps_1^\mu\veps_2^\nu
\\ & {} \times \biggl[ \ri k^\rho
  G_{\mu\nu\rho}^{\Pg^a\Pg^b\PZ}(p_1,p_2,-k) \, \frac{\xi\MZ}{\hat s} -
  G_{\mu\nu}^{\Pg^a\Pg^b\PG}(p_1,p_2,-k) \biggr],
\label{eq:M1PIb}
\end{split}
\end{equation}
where we have used $k=p_\PZ+p_\PH$, $p_\PZ\cdot\veps_{\PZ}^*=0$, and $k^2=\hat s$.
The terms in square brackets can be simplified using the well-known form
of the {\abbrev ABJ} anomaly.

To this end, we recall that the {\abbrev ABJ} anomaly relation for the axial
current $j_{f,5}^\mu=\overline\psi_f\gamma^\mu\gamma_5\psi_f$ and
the pseudo-scalar operator
$p_{f,5}=\overline\psi_f\gamma_5\psi_f$ 
for a quark $q$ reads
\beq
\partial_\mu j_{q,5}^\mu = 2\ri m_qp_{q,5}
-\frac{\alphas}{4\pi}F^a_{\mu\nu}\tilde F^{a,\mu\nu},
\label{eq:anomaly}
\eeq
where $\tilde F^{a,\mu\nu} = \frac{1}{2}\eps^{\mu\nu\rho\si} F^a_{\rho\si}$ is the dual
of the gluonic field-strength tensor $F^{a,\mu\nu}$.
\refeq{eq:anomaly} is an operator relation, valid for
bare quantities, expressing chiral symmetry. 
The Adler--Bardeen theorem~\cite{Adler:1969er}
states that it is correct to all orders in regularization schemes
that respect chiral symmetry. In regularizations that are not
chirally symmetric, \refeq{eq:anomaly} has to be restored
by extra counterterms from evanescent operators.
For the 't~Hooft--Veltman $\ga_5$ 
scheme~\cite{'tHooft:1972fi,Breitenlohner:1977hr}, which is employed in
our calculation, these are the counterterms
$\delta Z_5^{A}$ and 
$\delta Z_5^{P}$ calculated in
\citere{Larin:1993tq} and already used in \refse{sec:details}.
The relevant momentum-space Feynman rules for the 
composite operators $j_{q,5}^\mu$, $p_{q,5}$, and $F\tilde F$ 
are given by
\beqar
\includegraphics[width=.95\textwidth
,bb=75 690 540 800,clip
]{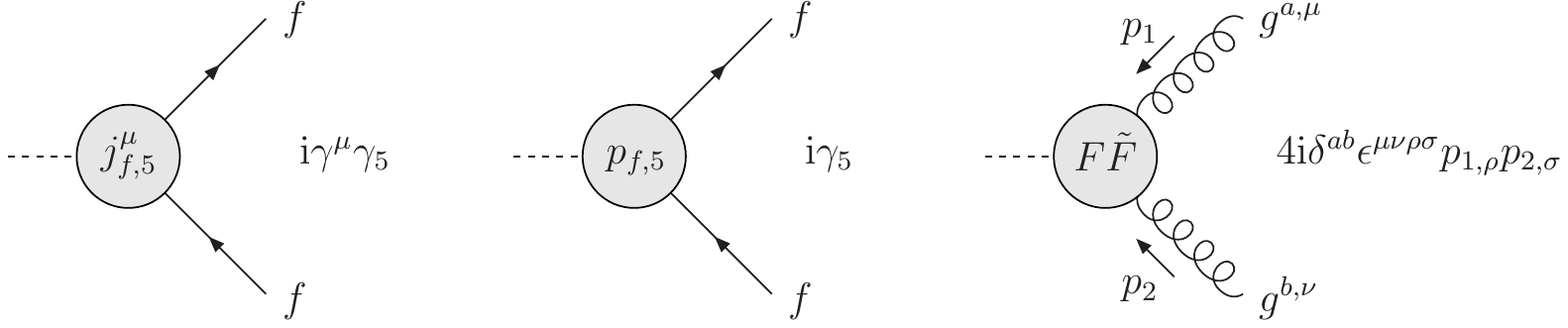}
\hspace*{-3em}
\nn
\\[-2em]
\eeqar
The dotted line in the Feynman rules indicate that the momentum
$p$ flows into the vertex.
The operator $F\tilde F$ induce Feynman rules with three and four
gluon lines as well, but those will not contribute in the following.
The fermionic Feynman rules are related to the couplings of
$\PZ/\PG$ to the quark:
\beq
Z^\mu\bar qq: -\frac{\ri eI^3_{\rw,q}}{2\sw\cw} \,\ga^\mu\ga_5 \,+\, \mbox{vector part,}
\qquad
\PG\bar qq: -\frac{eI^3_{\rw,q}}{\sw} \,\ga_5,
\eeq
where $I^3_{\rw,q}$ is the third component of the weak isospin of $q$.

Now we can make contact with the Green functions $G^{\Pg^a\Pg^b\PZ}$ and
$G^{\Pg^a\Pg^b\PG}$ introduced above. Since we consider only \qcd{}
corrections, in the relevant graphs the couplings of the external
$\PZ/\PG$ lines to an internal quark line represent the only electroweak
coupling in the $\Pg\Pg\PZ/\PG$ vertex functions.  These electroweak
couplings can be interpreted as the insertions of the operators
$j_{q,5}^\mu$ and $p_{q,5}$, because the vector part of the
$\PZ$~coupling does not contribute, as explained in
\refse{sec:outline}. 
Thus, we obtain
\begin{equation}
\begin{split}
G_{\mu\nu\rho}^{\Pg^a\Pg^b\PZ} &= \sum_q
G_{\mu\nu\rho}^{\Pg^a\Pg^b\PZ}\Big|_q, \qquad 
G_{\mu\nu\rho}^{\Pg^a\Pg^b\PZ}\Big|_q =
-\frac{\ri eI^3_{\rw,q}}{2\sw\cw} \, G_{\mu\nu\rho}^{\Pg^a\Pg^bj_{q,5}}, \qquad
\\
G_{\mu\nu}^{\Pg^a\Pg^b\PG} &= \sum_q
G_{\mu\nu}^{\Pg^a\Pg^b\PG}\Big|_q, \qquad
G_{\mu\nu}^{\Pg^a\Pg^b\PG}\Big|_q =
-\frac{eI^3_{\rw,q}}{\sw\cw} \, \frac{m_q}{\MZ} G_{\mu\nu}^{\Pg^a\Pg^bp_{q,5}}, 
\end{split}
\end{equation}
which is valid up to \nlo{} \qcd{},
where $G^{\Pg^a\Pg^b\PZ/\PG}|_q$ denotes the contributions induced 
by closed loops with quark $\Pq$.
Using this relation, \refeq{eq:anomaly} implies:
\beq
\ri p^\rho \, G_{\mu\nu\rho}^{\Pg^a\Pg^b\PZ}(p_1,p_2,p)\Big|_q =
-\MZ \, G_{\mu\nu}^{\Pg^a\Pg^b\PG}(p_1,p_2,p)\Big|_q 
+\frac{\ri eI^3_{\rw,q}}{2\sw\cw} \,
\frac{\alphas}{4\pi} \, G_{\mu\nu}^{\Pg^a\Pg^b(F\tilde F)}(p_1,p_2,p),
\eeq
where we have restored the arguments of the incoming momenta. 
Keeping in mind that $p=-k$, we thus can write the contribution of closed $q$-loops 
to $\M_n^{\mathrm{1PI}}(\eps)$ in \refeq{eq:M1PIb} in two different ways:
\beqar
\M_n^{\mathrm{1PI}}(\eps)|_q &=&
\frac{e}{\sw\cw} \,\frac{(p_\PH\cdot\veps_{\PZ}^*)}{\hat s} \,
\veps_1^\mu\veps_2^\nu
\nn\\
&& {} \times \biggl[
G_{\mu\nu}^{\Pg^a\Pg^b\PG}(p_1,p_2,-k)\Big|_q
+ \frac{\ri eI^3_{\rw,q}}{2\sw\cw} \,
\frac{\alphas}{4\pi} \, 
\frac{\xi\MZ}{\hat s-\xi\MZ^2}\, 
G_{\mu\nu}^{\Pg^a\Pg^b(F\tilde F)}(p_1,p_2,-k)
   \biggr],
\nn\\
&=& \frac{e}{\sw\cw} \,\frac{(p_\PH\cdot\veps_{\PZ}^*)}{\MZ \hat s} \,
\veps_1^\mu\veps_2^\nu
\label{eq:M1PIc}
\\
&& {} \times \biggl[
\ri k^\rho G_{\mu\nu\rho}^{\Pg^a\Pg^b\PZ}(p_1,p_2,-k)\Big|_q
+\frac{\ri eI^3_{\rw,q}}{2\sw\cw} \,
\frac{\alphas}{4\pi} \, \frac{\hat s}{\hat s-\xi\MZ^2} \,
G_{\mu\nu}^{\Pg^a\Pg^b(F\tilde F)}(p_1,p_2,-k)
   \biggr].
\nn
\eeqar
Now we exploit our approximation of $\Mt\to\infty$ and $m_{q\ne\Pt}=0$,
which implies
$k^\rho G^{\Pg^a\Pg^b\PZ}|_\Pt=0$
and 
$G^{\Pg^a\Pg^b\PG}|_{q\ne\Pt}=0$.
The former result is taken from our diagrammatical large-mass expansion, the
latter is a trivial consequence of the vanishing Yukawa couplings of the
massless quarks.
For the top quark we take the second form of the last equation 
of \refeq{eq:M1PIc} and for the other
quarks the first, and obtain
\begin{equation}
\begin{split}
\M_n^{\mathrm{1PI}}(\eps) &=
\frac{e}{\sw\cw} \,\frac{(p_\PH\cdot\veps_{\PZ}^*)}{\hat s} \,
\veps_1^\mu\veps_2^\nu
\biggl[
\sum_{q\ne\Pt}
\frac{\ri eI^3_{\rw,q}}{2\sw\cw} \,
\frac{\alphas}{4\pi} \, 
\frac{\xi\MZ}{\hat s-\xi\MZ^2}\, 
G_{\mu\nu}^{\Pg^a\Pg^b(F\tilde F)}(p_1,p_2,-k)
\\
& \qquad {}
+\frac{\ri eI^3_{\rw,\Pt}}{2\sw\cw} \,
\frac{\alphas}{4\pi} \, \frac{\hat s}{\hat s-\xi\MZ^2} \,
G_{\mu\nu}^{\Pg^a\Pg^b(F\tilde F)}(p_1,p_2,-k)
   \biggr]
\\
&= 
\frac{\ri\alpha\alphas}{4\sw^2\cw^2} \,
\frac{(p_\PH\cdot\veps_{\PZ}^*)}{\MZ \hat s} \,
\veps_1^\mu\veps_2^\nu \,
G_{\mu\nu}^{\Pg^a\Pg^b(F\tilde F)}(p_1,p_2,-k),
\end{split}
\end{equation}
where the dependence on the gauge parameter $\xi$ cancels, as it should be.
For the \lo{} matrix element, the Green function 
$G^{\Pg^a\Pg^b(F\tilde F)}$ just has to be replaced by the
Feynman rule for the $F\tilde F$ operator with two external gluon legs,
yielding
\beq
\M_0 = 
-\delta^{ab}\,
\frac{\alpha\alphas}{\sw^2\cw^2\MZ} \,
\frac{(p_\PH\cdot\veps_{\PZ}^*)}{\hat s} \,
\eps(\veps_1,\veps_2,p_1,p_2)
\label{eq:M0}
\eeq
in agreement with \refeq{eq:M0square}.
The \nlo{} \qcd{} corrections to $G^{\Pg^a\Pg^b(F\tilde F)}$
are induced by the diagrams shown in \reffi{fig:FF}. 
\bfi
\centerline{
\includegraphics[width=.95\textwidth
,bb=80 570 490 800,clip
]{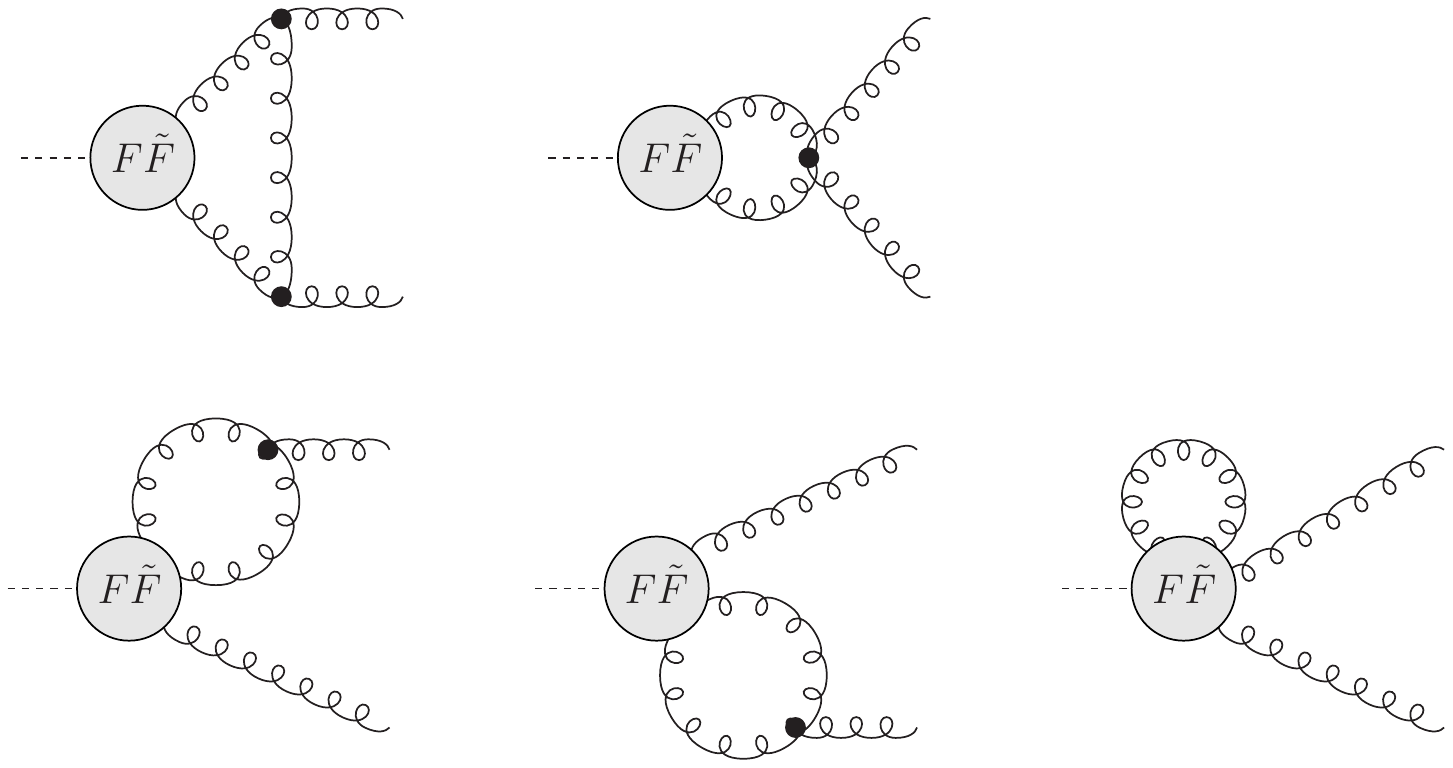}
}
\caption{\nlo{} \qcd{} diagrams for the Green function 
$G^{\Pg^a\Pg^b(F\tilde F)}$.}
\label{fig:FF}
\efi
The actual calculation of these one-loop diagrams is very simple.
For on-shell gluons only the first two diagrams contribute and
yield
\beq
\M_1^{\mathrm{1PI,virt}}(\eps) =
\frac{3\alphas}{2\pi}\,\Bigl[ 2-\hat s C_0(\hat s,0,0,0,0,0,0) \Bigr] \,\M_0,
\eeq
where $C_0$ is the one-loop scalar 3-point integral in the convention of
\citeres{Denner:2002ii,Denner:2005nn,Denner:2010tr} (see e.g.\ 
\citere{Dittmaier:2003bc} for the explicit result).
This is the unrenormalized result for the virtual correction which
still has to be renormalized. As stated above, the anomaly
equation (\ref{eq:anomaly}) is valid for bare quantities, so that
the $F\tilde F$ term receives renormalization contributions from
$\alphas$ and the gluon field. Since in our case only the 
two-gluon contribution of $F\tilde F$ is relevant, we get the
simple factorizing contribution to the \nlo{} amplitude,
\beq
\M_1^{\mathrm{1PI,ct}}(\eps) =
(2\delta Z_g+\delta Z_3) \,\M_0.
\eeq
Combining the renormalized virtual amplitude 
$\M_1^{\mathrm{1PI,virt}}+\M_1^{\mathrm{1PI,ct}}$ with the
contribution $\delta_\text{CS}$, see \refeq{eq:deltaCS},
of the Catani--Seymour $I$-operator
of the subtraction function yields the 1PI part
of the correction to the cross section, as given in
\refeq{eq:sigma-virt} by first term on the r.h.s.

\section*{Acknowledgements}
We would like to thank O.\,Brein and W.\,Kilgore for useful
communication.  This work was supported by {\abbrev DFG}, contract
HA~2990/5-1, the Helmholtz Alliance ``Physics at the Terascale''.

\def\app#1#2#3{{\it Act.~Phys.~Pol.~}\jref{\bf B #1}{#2}{#3}}
\def\apa#1#2#3{{\it Act.~Phys.~Austr.~}\jref{\bf#1}{#2}{#3}}
\def\annphys#1#2#3{{\it Ann.~Phys.~}\jref{\bf #1}{#2}{#3}}
\def\cmp#1#2#3{{\it Comm.~Math.~Phys.~}\jref{\bf #1}{#2}{#3}}
\def\cpc#1#2#3{{\it Comp.~Phys.~Commun.~}\jref{\bf #1}{#2}{#3}}
\def\epjc#1#2#3{{\it Eur.\ Phys.\ J.\ }\jref{\bf C #1}{#2}{#3}}
\def\fortp#1#2#3{{\it Fortschr.~Phys.~}\jref{\bf#1}{#2}{#3}}
\def\ijmpc#1#2#3{{\it Int.~J.~Mod.~Phys.~}\jref{\bf C #1}{#2}{#3}}
\def\ijmpa#1#2#3{{\it Int.~J.~Mod.~Phys.~}\jref{\bf A #1}{#2}{#3}}
\def\jcp#1#2#3{{\it J.~Comp.~Phys.~}\jref{\bf #1}{#2}{#3}}
\def\jetp#1#2#3{{\it JETP~Lett.~}\jref{\bf #1}{#2}{#3}}
\def\jphysg#1#2#3{{\small\it J.~Phys.~G~}\jref{\bf #1}{#2}{#3}}
\def\jhep#1#2#3{{\small\it JHEP~}\jref{\bf #1}{#2}{#3}}
\def\mpl#1#2#3{{\it Mod.~Phys.~Lett.~}\jref{\bf A #1}{#2}{#3}}
\def\nima#1#2#3{{\it Nucl.~Inst.~Meth.~}\jref{\bf A #1}{#2}{#3}}
\def\npb#1#2#3{{\it Nucl.~Phys.~}\jref{\bf B #1}{#2}{#3}}
\def\nca#1#2#3{{\it Nuovo~Cim.~}\jref{\bf #1A}{#2}{#3}}
\def\plb#1#2#3{{\it Phys.~Lett.~}\jref{\bf B #1}{#2}{#3}}
\def\prc#1#2#3{{\it Phys.~Reports }\jref{\bf #1}{#2}{#3}}
\def\prd#1#2#3{{\it Phys.~Rev.~}\jref{\bf D #1}{#2}{#3}}
\def\pR#1#2#3{{\it Phys.~Rev.~}\jref{\bf #1}{#2}{#3}}
\def\prl#1#2#3{{\it Phys.~Rev.~Lett.~}\jref{\bf #1}{#2}{#3}}
\def\pr#1#2#3{{\it Phys.~Reports }\jref{\bf #1}{#2}{#3}}
\def\ptp#1#2#3{{\it Prog.~Theor.~Phys.~}\jref{\bf #1}{#2}{#3}}
\def\ppnp#1#2#3{{\it Prog.~Part.~Nucl.~Phys.~}\jref{\bf #1}{#2}{#3}}
\def\rmp#1#2#3{{\it Rev.~Mod.~Phys.~}\jref{\bf #1}{#2}{#3}}
\def\sovnp#1#2#3{{\it Sov.~J.~Nucl.~Phys.~}\jref{\bf #1}{#2}{#3}}
\def\sovus#1#2#3{{\it Sov.~Phys.~Usp.~}\jref{\bf #1}{#2}{#3}}
\def\tmf#1#2#3{{\it Teor.~Mat.~Fiz.~}\jref{\bf #1}{#2}{#3}}
\def\tmp#1#2#3{{\it Theor.~Math.~Phys.~}\jref{\bf #1}{#2}{#3}}
\def\yadfiz#1#2#3{{\it Yad.~Fiz.~}\jref{\bf #1}{#2}{#3}}
\def\zpc#1#2#3{{\it Z.~Phys.~}\jref{\bf C #1}{#2}{#3}}
\def\ibid#1#2#3{{ibid.~}\jref{\bf #1}{#2}{#3}}
\def\otherjournal#1#2#3#4{{\it #1}\jref{\bf #2}{#3}{#4}}
\newcommand{\jref}[3]{{\bf #1} (#2) #3}
\newcommand{\hepph}[1]{\href{http://arxiv.org/abs/hep-ph/#1}{\tt [hep-ph/#1]}}
\newcommand{\mathph}[1]{\href{http://arxiv.org/abs/math-ph/#1}{\tt
    [math-ph/#1]}}
\newcommand{\hepth}[1]{\href{http://arxiv.org/abs/hep-th/#1}{\tt [hep-th/#1]}}
\newcommand{\arxiv}[2]{\href{http://arxiv.org/abs/#1}{\tt [arXiv:#1]}}
\newcommand{\bibentry}[4]{#1, ``#2,'' #3\ifthenelse{\equal{#4}{}}{}{, }#4.}


\begin{thebibliography}{99}
%
%

\bibitem{Aad:2012gk}
  \bibentry{G.~Aad {\it et al.} [ATLAS Collaboration]} {Observation of
  a new particle in the search for the Standard Model Higgs boson with
  the ATLAS detector at the LHC} {\plb{716}{2012}{1}} 
 {\arxiv{1207.7214}{hep-ex}}
%
%

\bibitem{Chatrchyan:2012gu}
  \bibentry{S.~Chatrchyan {\it et al.}  [CMS Collaboration]}
           {Observation of a new boson at a mass of 125 GeV with the CMS
             experiment at the LHC} 
           {\plb{716}{2012}{30}}
           {\arxiv{1207.7235}{hep-ex}}
%
%

\bibitem{Aaltonen:2012qt}
  \bibentry{T.~Aaltonen {\it et al.}  [CDF and D0 Collaborations]}
  {Evidence for a particle produced in association with weak bosons and decaying to a bottom-antibottom quark pair in Higgs boson searches at the Tevatron}
  {\prl{109}{2012}{071804}}
  {\arxiv{1207.6436}{hep-ex}}
%
%

\bibitem{Dittmaier:2011ti}
  \bibentry{S.~Dittmaier {\it et al.} [LHC Higgs Cross Section Working
  Group Collaboration]}
  {Handbook of LHC Higgs cross sections: 1. inclusive observables}
  {\arxiv{1101.0593}{hep-ph}}
  {}
%
%

\bibitem{Dittmaier:2012vm}
  \bibentry{S.~Dittmaier {\it et al.}  [LHC Higgs Cross Section Working
  Group Collaboration]} {Handbook of LHC Higgs cross sections:
  2. differential distributions} {\arxiv{1201.3084}{hep-ph}}{}
%
%

\bibitem{Butterworth:2008iy}
  \bibentry{J.M.~Butterworth, A.R.~Davison, M.~Rubin, and G.P.~Salam}
  {Jet substructure as a new Higgs search channel at the LHC}
  {\prl{100}{2008}{242001}}
  {\arxiv{0802.2470}{hep-ph}}
%
%

\bibitem{Hamberg:1991np}
\bibentry{R.~Hamberg, W.L.~van Neerven, and T.~Matsuura}
{A complete calculation of the order $\alpha_s^2$ correction to the
Drell-Yan K factor}
{\npb{359}{1991}{343}, (E)~\ibid{B 644}{2002}{403}}
{}
%
%

\bibitem{Harlander:2002wh}
\bibentry{R.V.~Harlander and W.B.~Kilgore}
{Next-to-next-to-leading order Higgs production at hadron colliders}
{\prl{88}{2002}{201801}}
{\hepph{0201206}}

\bibitem{Brein:2003wg}
\bibentry{O.~Brein, A.~Djouadi, and R.~Harlander}
{{\abbrev NNLO} QCD corrections to the Higgs-strahlung
processes at hadron colliders}
{\plb{579}{2004}{149}}
{\hepph{0307206}}
%
%

\bibitem{Ferrera:2011bk}
  \bibentry{G.~Ferrera, M.~Grazzini, and F.~Tramontano} {Associated $\PW\PH$
  production at hadron colliders: a fully exclusive QCD calculation at
  NNLO} 
  {\prl{107}{2011}{152003}}
  {\arxiv{1107.1164}{hep-ph}}
%
%

\bibitem{Brein:2011vx}
  \bibentry{O.~Brein, R.~Harlander, M.~Wiesemann, and T.~Zirke}
  {Top-Quark mediated effects in hadronic Higgs-Strahlung}
  {\arxiv{1111.0761}{hep-ph}}
  {\epjc{72}{2012}{1868}}
%
%

\bibitem{Ciccolini:2003jy}
\bibentry{M.L.~Ciccolini, S.~Dittmaier, and M.~Kr\"amer}
{Electroweak radiative corrections to associated $\PW\PH$ and $\PZ\PH$ production  at
hadron colliders} 
{\prd{68}{2003}{073003}}
{\hepph{0306234}}
%
%

\bibitem{Denner:2011id}
  \bibentry{A.~Denner, S.~Dittmaier, S.~Kallweit, and A.~M\"uck}
  {Electroweak corrections to Higgs-strahlung off $\PW/\PZ$ bosons at the
  Tevatron and the LHC with HAWK} 
  {\jhep{1203}{2012}{075}}
  {\arxiv{1112.5142}{hep-ph}}
%
%

\bibitem{Brein:2004ue}
\bibentry{O.~Brein, M.~Ciccolini, S.~Dittmaier, A.~Djouadi, 
R.~Harlander, and M.~Kr\"amer}
{Precision calculations for associated $\PW\PH$ and $\PZ\PH$ production at hadron
colliders}
{\hepph{0402003}, in:  K.A.~Assamagan {\it et al.},
\hepph{0406152}}
{}
%
%

\bibitem{Banfi:2012jh}
  \bibentry{A.~Banfi, J.~Cancino, and J.~Cancino}
  {Implications of QCD radiative corrections on high-$p_T$ Higgs searches}
  {\arxiv{1207.0674}{hep-ph}}
   {}
%
%

\bibitem{Dawson:1998py}
\bibentry{S.~Dawson, S.~Dittmaier, and M.~Spira}
{Neutral Higgs-boson pair production at hadron colliders: QCD corrections}
{\prd{58}{1998}{115012}}
{\hepph{9805244}}
%
%

\bibitem{landau}
  L.D.~Landau,
  {\it Dokl.\ Akad.\ Nawk.} {\bf 60} (1948) 207.
%
%

\bibitem{Yang:1950rg}
  \bibentry{C.-N.~Yang}
  {Selection rules for the dematerialization of a particle into two photons}
  {\pR{77}{1950}{242}}
   {}
%
%
%

\bibitem{Kniehl:1990iv}
\bibentry{B.A.~Kniehl}
{Associated production of Higgs and $Z$ bosons from gluon fusion in hadron
collisions}
{\prd{42}{2253}{1990}}
{}
%
%
%
%

\bibitem{vhnnlo}
\bibentry{O.~Brein, R.V.~Harlander, and T.J.E.~Zirke}
{{\tt vh@nnlo}: a program to evaluate hadronic Higgs-Strahlung at
next-to-next-to-leading order}
{\arxiv{1210.5347}{hep-ph}}
{}
%
%

\bibitem{Spira:1995rr}
\bibentry{M.~Spira, A.~Djouadi, D.~Graudenz, and P.M.~Zerwas}
{Higgs boson production at the {\abbrev LHC}}
{\npb{453}{1995}{17}}
{\hepph{9504378}}
%
%

\bibitem{Kramer:1996iq}
\bibentry{M.~Kr\"amer, E.~Laenen, and M.~Spira}
{Soft gluon radiation in Higgs boson production at the {\abbrev LHC}}
{\npb{511}{1998}{523}}
{\hepph{9611272}}
%
%

\bibitem{Harlander:2010my}
  \bibentry{R.V.~Harlander, H.~Mantler, S.~Marzani, and K.J.~Ozeren}
  {Higgs production in gluon fusion at next-to-next-to-leading order QCD for
  finite top mass}
  {\epjc{66}{2010}{359}}
  {\arxiv{0912.2104}{hep-ph}}
%
%

\bibitem{Harlander:2009bw}
\bibentry{R.V.~Harlander and K.J.~Ozeren}
{Top mass effects in Higgs production at next-to-next-to-leading order QCD:
virtual corrections}
{\plb{679}{2009}{467}}
{\arxiv{0907.2997}{hep-ph}}
%
%

\bibitem{Pak:2009dg}
\bibentry{A.~Pak, M.~Rogal, and M.~Steinhauser}
{Finite top quark mass effects in NNLO Higgs boson production at LHC}
{\jhep{1002}{2010}{025}}
{\arxiv{0911.4662}{hep-ph}}
%
%

\bibitem{Pak:2011hs}
  \bibentry{A.~Pak, M.~Rogal, and M.~Steinhauser}
  {Production of scalar and pseudo-scalar Higgs bosons to
  next-to-next-to-leading order at hadron colliders}
  {\jhep{1109}{2011}{088}}
  {\arxiv{1107.3391}{hep-ph}}
%
%

\bibitem{Harlander:2012hf}
  \bibentry{R.V.~Harlander, T.~Neumann, K.J.~Ozeren, and M.~Wiesemann}
  {Top-mass effects in differential Higgs production through gluon fusion at order $\alpha_s^4$}
  {\jhep{1208}{2012}{139}}
  {\arxiv{1206.0157}{hep-ph}}
%
%

\bibitem{Smirnov:2002pj}
\bibentry{V.A.~Smirnov}
{Applied asymptotic expansions in momenta and masses}
{\otherjournal{Springer Tracts Mod.\ Phys.\ }{177}{2002}{1}}
{}
%
%

\bibitem{Smirnov:1994tg}
\bibentry{V.A.~Smirnov}
{Asymptotic expansions in momenta and masses and calculation of Feynman
diagrams}
{\mpl{10}{1995}{1485}}
{\hepth{9412063}}
%
%

\bibitem{'tHooft:1972fi}
  \bibentry{G.~'t Hooft and M.J.G.~Veltman}
  {Regularization and renormalization of gauge fields}
  {\npb{44}{1972}{189}}
  {}
%
%

\bibitem{Breitenlohner:1977hr}
\bibentry{P.~Breitenlohner and D.~Maison}
{Dimensional renormalization and the action principle}
{\cpc{52}{1977}{11}}
{}
%
%

\bibitem{Larin:1993tq}
\bibentry{S.A.~Larin}
{The renormalization of the axial anomaly in dimensional regularization}
{\plb{303}{1993}{113}}
{\hepph{9302240}}
%
%

\bibitem{Kublbeck:1990xc}
\bibentry{J.~K\"ublbeck, M.~B\"ohm, and A.~Denner}
{FeynArts: Computer algebraic generation of Feynman graphs and
amplitudes}
{\cpc{60}{1990}{165}}
{}
%
%

\bibitem{Nogueira:1991ex}
\bibentry{P.~Nogueira}
{Automatic Feynman graph generation}
{\jcp{105}{1993}{279}}
{}
%
%
%
%

\bibitem{Harlander:1997zb}
\bibentry{R.~Harlander, T.~Seidensticker, and M.~Steinhauser}
{Corrections of ${\cal O}(\alpha \alpha_s)$ to the decay of the $Z$ boson  
into bottom quarks}
{\plb{426}{1998}{125}}
{\hepph{9712228}}
%
%

\bibitem{Seidensticker:1999bb}
\bibentry{T.~Seidensticker}
{Automatic application of successive asymptotic expansions of Feynman
diagrams}
{\hepph{9905298}}
{}
%
%

\bibitem{Steinhauser:2000ry}
\bibentry{M.~Steinhauser}
{{\tt MATAD}: a program package for the computation of massive tadpoles}
{\cpc{134}{2001}{335}}
{\hepph{0009029}}
%
%

\bibitem{Harlander:2000mg}
\bibentry{R.V.~Harlander}
{Virtual corrections to $\Pg \Pg \to \PH$ to two loops in the heavy top limit}
{\plb{492}{2000}{74}}
{\hepph{0007289}}
%
%

\bibitem{Catani:1996vz}
\bibentry{S.~Catani and M.H.~Seymour}
{A general algorithm for calculating jet cross sections in NLO QCD}
{\npb{485}{1997}{291}; (E) \ibid{510}{1998}{503}}
{\hepph{9605323}}
%
%

\bibitem{Chetyrkin:2000yt}
  \bibentry{K.G.~Chetyrkin, J.H.~K\"uhn, and M.~Steinhauser}
  {RunDec: a Mathematica package for running and decoupling of the
  strong coupling and quark masses}
  {\cpc{133}{2000}{43}}
  {\hepph{0004189}}
%
%

\bibitem{'tHooft:1978xw}
\bibentry{G.~'t Hooft and M.J.G.~Veltman}
{Scalar one loop integrals}
{\npb{153}{1979}{365}}
{}
%
%

\bibitem{Passarino:1978jh}
\bibentry{G.~Passarino and M.J.G.~Veltman}
{One loop corrections for $e^+ e^-$ annihilation into $\mu^+ \mu^-$ in
the Weinberg model}
{\npb{160}{1979}{151}} {}
%
%

\bibitem{Beenakker:1988jr}
  \bibentry{W.~Beenakker and A.~Denner}
  {Infrared divergent scalar box integrals with applications in the
  electroweak Standard Model}
  {\npb{338}{1990}{349}}
  {}
%
%

\bibitem{Denner:2002ii}
\bibentry{A.~Denner and S.~Dittmaier}
{Reduction of one-loop tensor 5-point integrals}
{\npb{658}{2003}{175}}
{\hepph{0212259}}
%
%

\bibitem{Denner:2005nn}
\bibentry{A.~Denner and S.~Dittmaier}
{Reduction schemes for one-loop tensor integrals}
{\npb{734}{2006}{62}}
{\hepph{0509141}}
%
%

\bibitem{Denner:2010tr}
  \bibentry{A.~Denner and S.~Dittmaier}
  {Scalar one-loop 4-point integrals}
  {\npb{844}{2011}{199}}
  {\arxiv{1005.2076}{hep-ph}}
%
%

\bibitem{Hahn:2000kx}
\bibentry{T.~Hahn}
{Generating Feynman diagrams and amplitudes with FeynArts 3}
{\cpc{140}{2001}{418}}
{\hepph{0012260}}
%
%

\bibitem{Hahn:1998yk}
\bibentry{T.~Hahn and M.~Perez-Victoria}
{Automatized one-loop calculations in four and $D$ dimensions}
{\cpc{118}{1999}{153}}
{\hepph{9807565}}
%
%

\bibitem{Hahn:2000jm}
  \bibentry{T.~Hahn}
  {Automatic loop calculations with FeynArts, FormCalc, and LoopTools}
  {{\it Nucl.\ Phys.\ Proc.\ Suppl.\ } {\bf 89} (2000) 231}
  {\hepph{0005029}}
%
%

\bibitem{Vermaseren:2000nd}
\bibentry{J.A.~Vermaseren}
{New features of {\abbrev FORM}}
{\mathph{0010025}}
{}
%
%
%
%

\bibitem{Ellis:2007qk}
  \bibentry{R.~K.~Ellis and G.~Zanderighi}
  {Scalar one-loop integrals for QCD}
  {\jhep{0802}{2008}{002}}
  {\arxiv{0712.1851}{hep-ph}}
%
%

\bibitem{Martin:2009iq}
  \bibentry{A.D.~Martin, W.J.~Stirling, R.S.~Thorne, and G.~Watt}
  {Parton distributions for the LHC}
  {\epjc{63}{2009}{189}}
  {\arxiv{0901.0002}{hep-ph}}
%
%

\bibitem{lhapdf}
 \bibentry{M.R.~Whalley, D.~Bourilkov, and R.C.~Group}
  {The Les Houches accord PDFs (LHAPDF) and LHAGLUE}
  {\hepph{0508110}}
  {\href{http://projects.hepforge.org/lhapdf/}
  {\tt http://projects.hepforge.org/lhapdf/}}
%
%

\bibitem{Botje:2011sn}
  \bibentry{M.~Botje {\it et al.}}
  {The PDF4LHC Working Group Interim Recommendations}
  {\arxiv{1101.0538}{hep-ph}}
   {}
%
%

\bibitem{Lai:2010vv}
  \bibentry{H.-L.~Lai {\it et al.}}
  {New parton distributions for collider physics}
  {\prd{82}{2010}{074024}}
  {\arxiv{1007.2241}{hep-ph}}
%
%
%
%

\bibitem{Ball:2012cx}
  \bibentry{R.D.~Ball, V.~Bertone, S.~Carrazza, C.S.~Deans, L.~Del
  Debbio, S.~Forte, A.~Guffanti, N.P.~Hartland {\it et al.}}
  {Parton distributions with LHC data}
  {\arxiv{1207.1303}{hep-ph}}
   {}
%
%

\bibitem{Adler:1969gk}
  \bibentry{S.L.~Adler}
  {Axial vector vertex in spinor electrodynamics}
  {\pR{177}{1969}{2426}}
  {}
%
%

\bibitem{Adler:1969er}
  \bibentry{S.L.~Adler and W.A.~Bardeen}
  {Absence of higher order corrections in the anomalous axial vector
  divergence equation}
  {\pR{182}{1969}{1517}}
  {}
%
%

\bibitem{Denner:1991kt}
\bibentry{A.~Denner}
{Techniques for calculation of electroweak radiative corrections at the one
 loop level and results for W physics at LEP-200}
{\fortp{41}{1993}{307}}
{}
%
%

\bibitem{Dittmaier:2003bc}
  \bibentry{S.~Dittmaier}
  {Separation of soft and collinear singularities from one loop $N$ point
  integrals}
  {\npb{675}{2003}{447}}
  {\hepph{0308246}}
%
%

\end{thebibliography}
\end{document}